\documentclass[useAMS,usenatbib,usegraphicx]{mn2e}

\title[Component separation and non-Gaussianity]
{Effect of component separation on the temperature distribution of the CMB}
\author[Barreiro et al.]
{R.B.~Barreiro$^1$, E.~Mart\'\i nez-Gonz\'alez$^1$, P.~Vielva$^{1,2}$ and
M.P.~Hobson$^3$ \\ 
$^1$Instituto de F\'\i sica de Cantabria, CSIC-Universidad de Cantabria,
Santander, Spain \\
$^2$ PCC - Coll\`ege de France, 11, place Marcelin Berthelot, F-75231
Paris, France \\ 
$^3$Astrophysics Group, Cavendish Laboratory, Madingley Road,
Cambridge CB3 0HE \\}
\date{Accepted ---. Received ---; in original form \today}
\pubyear{2005}

\begin{document}
\maketitle
\begin{abstract}
We present a study of the effect of component separation on the
recovered cosmic microwave background (CMB) temperature
distribution, considering Gaussian and non-Gaussian input CMB maps. In
particular, the non-Gaussian maps have been generated as a mixture of
a Gaussian CMB map and a cosmic strings map in different proportions.
First, we extract the CMB component from simulated multifrequency
Planck data (in small patches of the sky) using the maximum-entropy
method (MEM), Wiener filter (WF) and a method based on the subtraction
of foreground templates plus a linear combination of frequency
channels (LCFC).  We then apply a wavelet-based method to study the
Gaussianity of the recovered CMB and compare it with the same analysis
for the input map.  When the original CMB map is Gaussian (and
assuming that point sources have been removed), we find that neither
MEM nor WF introduce non-Gaussianity in the CMB
reconstruction. Regarding the LCFC, the Gaussian character is also
preserved provided that the appropriate combination of frequency
channels is used. On the contrary, if the input CMB map is
non-Gaussian, all the studied methods produce a reconstructed CMB with
lower detections of non-Gaussianity than the original map. This effect
is mainly due to the presence of instrumental noise in the data, which
clearly affects the quality of the reconstructions.  In this case, MEM
tends to produce slightly higher non-Gaussian detections in the
reconstructed map than WF whereas the detections are lower for the
LCFC. We have also studied the effect of point sources in the MEM
reconstruction. If no attempt to remove point sources is performed,
they clearly contaminate the CMB reconstruction, introducing spurious
non-Gaussianity. When the brightest point sources are removed from the
data using the Mexican Hat Wavelet, the Gaussian character of the CMB
is preserved. However, when analysing larger
regions of the sky, the variance of our estimators will be
appreciably reduced and, in this case, we expect the point source
residuals to introduce spurious non-Gaussianity in the CMB
distribution. Therefore a careful subtraction (or masking) of point
source emission is crucial in order to be able to perform Gaussian
analysis of the CMB.

\end{abstract}


\section{Introduction}
One of the most valuable tools of modern cosmology is the observation
and analysis of anisotropies in the cosmic microwave background (CMB)
radiation.  A number of experiments such as Boomerang~\citep{net02},
MAXIMA~\citep{han00}, DASI~\citep{hal02}, VSA~\citep{gra03},
CBI~\citep{mas03}, ACBAR~\citep{kuo04}, Archeops~\citep{ben03} and
WMAP~\citep{ben03a} have already provided measurements of the
power spectrum of the CMB, allowing one to put tight constraints on the
cosmological parameters. In addition, the temperature distribution of
the CMB also carries crucial information about the theory of structure
formation. In particular, the standard inflationary theory predicts
Gaussian fluctuations for the CMB whereas topological defect models
(\citealt{tur90}, ~\citealt{dur99}) and non-standard inflation
(\citealt{lin97}, ~\citealt{pee97}, ~\citealt{ber02},
~\citealt{acq03}, ~\citealt{bart04}) introduce non-Gaussian signatures
on the cosmological signal. Therefore, a detection of intrinsic
non-Gaussianity in the CMB would have far reaching consequences for
the current understanding of cosmology.

When observing the microwave sky, however, one measures not only
the cosmological signal but a mixture of the CMB with other
contaminant components. The most important of these are the emissions
from our own Galaxy (mainly synchrotron, free-free and dust emission),
the thermal and kinematic Sunyaev-Zeldovich effects and the emission
from extragalactic point sources. In addition, the data will also be
corrupted by instrumental noise and possibly some systematic effects.
Therefore, in order to extract all the valuable information encoded in
the CMB, it is critical to separate the cosmological signal from the
other microwave components. This is especially relevant for the
success of future CMB experiments, which will measure the microwave
sky with unprecedented resolution, sensitivity, sky and frequency
coverage. Most notably, these include the WMAP mission by NASA (that
will continue to take data in the next few years) and the Planck mission
by ESA (to be launched in 2007), both of which produce all-sky
multifrequency observations of the CMB.  The problem of component
separation is a particularly important issue in the analysis of the
temperature distribution of the CMB, since foregrounds and/or
systematics may well introduce non-Gaussian signatures on the
cosmological signal or, conversely, impair our ability to detect
underlying intrinsic non-Gaussianity.  The component separation
process itself should also be well understood in order to avoid the
introduction of artifacts that can modify the temperature distribution
of the CMB.

The study of the Gaussianity of the CMB has recently attracted a great
interest with the release of the 1st-year WMAP data~\citep{ben03a}.
There have been many analyses of the temperature distribution of the
CMB
(\citealt{kom03},~\citealt{col03},~\citealt{gw03},~\citealt{gaz03},
~\citealt{chi03},~\citealt{eri04a,eri04b},~\citealt{par04},
~\citealt{mag04},~\citealt{vie04},~\citealt*{cop04},~\citealt{han04},
~\citealt{muk04}, ~\citealt{lar04}, ~\citealt{col04},
~\citealt{sch04}, ~\citealt{cab04},
~\citealt{chi04}, ~\citealt{cru05}, 
~\citealt{mce05}, ~\citealt{lan05}, ~\citealt*{cay05}) and, although
for some of them the data are consistent with Gaussian fluctuations ,
in other cases there has been a detection of non-Gaussianity and/or
asymmetries, whose origin can not always be attributed to the presence
of foregrounds or systematics.  For instance,~\cite{vie04} use a
method based on the spherical Mexican Hat wavelet (MHW) and claim a
non-Gaussian detection outside the 99 per cent acceptance region, for
which an intrinsic origin can not be discarded. By extending the
analysis method to orientable spherical wavelets, ~\cite{mce05}
also obtained a significant non-Gaussian detection.

With regard to the component separation problem, several methods have
been proposed in the literature. There are basically two different
approaches. The first one is to design a method to extract only a
particular component from the microwave sky. The second one attempts
to reconstruct all the components at the same time. The first
approach is particularly well suited to the extraction of compact
sources (for a review see e.g. \citealt{bar06}), such as the emission
from extragalactic point sources or the thermal and kinematic
Sunyaev-Zeldovich (SZ) effects. Such methods include the use of the
MHW (\citealt{cay00}, ~\citealt{vie01a,vie01b,vie03}) the matched
filter (\citealt{teg98}, ~\citealt{her02a}, ~\citealt{pie03},
~\citealt{sch03}, ~\citealt{lop04},~\citealt{her04},
~\citealt{scha04},~\citealt*{mel05}), the scale-adaptive filter
(\citealt*{san01}, ~\citealt{her02a,her02b,her02c}, \citealt{bar03}),
the adaptive top-hat filter~\citep{chi02}, the biparametric scale
adaptive filter~\citep{lop05,lop05b}, the Bayesian approach proposed
by~\cite{die02}, the McClean algorithm~\citep{hob03} or the
wavelet-based method of \cite{pie04}. In addition, there are methods
that try to extract only the CMB component from the data, such as the
blind EM algorithm of \cite{mar03} or the internal linear combination
method (\citealt{ben03b}, \citealt{teg03}, \citealt{eri04c}).
Regarding the techniques that attempt to separate and reconstruct all
the components at the same time, we have Wiener filter
(WF,~\citealt{bou96,teg96}), the maximum-entropy method
(MEM,~\citealt{hob98,hob99a},~\citealt{sto02,sto05},~\citealt{bar04},
~\citealt{ben03b}) or blind source separation
(\citealt{bac00},~\citealt{mai02,mai03}, ~\citealt*{del03},
~\citealt{bed04},~\citealt{pat04}).

The effect that these techniques have on the recovered power spectrum
of the CMB have been studied in many of the cases. However, no
attention has been paid to the effect that the component separation
may have on the underlying CMB temperature distribution. As already
mentioned, the Gaussianity of the CMB fluctuations is predicted by the
standard inflationary model and, therefore, a detection of intrinsic
non-Gaussianity would have a great impact on our current knowledge of
the universe.  Thus, a complete understanding of the processing of the
data is crucial since otherwise we could either misidentify spurious
non-Gaussian signatures as true ones or impair our ability to detect
intrinsic non-Gaussianity. In the present work, we will study the
effect of some component separation techniques on the underlying CMB
temperature distribution. In order to do this, first we will apply
different component separation techniques (MEM, WF, combined MEM+MHW
and a linear combination of the frequency channels, LCFC) to simulated
Planck data of small patches of the sky. We will perform then a
Gaussianity analysis based on a wavelet technique to the reconstructed
CMB map and compare the results with the same analysis for the input
map.

The outline of the paper is as follows. In Section~\ref{sec:cpt_sep}
we summarise the different component separation methods that are
applied in this work. The wavelet technique used for the Gaussian
analysis is explained in Section~\ref{sec:wavelets} whereas
Section~\ref{sec:simulations} describes the Planck simulated data. Our
results are given in Section~\ref{sec:results} and
section~\ref{sec:role} presents some additional tests to understand
further the role of foregrounds and noise in the component separation
process.  Finally a discussion and our conclusions are presented in
Sections~\ref{sec:discussion} and ~\ref{sec:conclusions} respectively.

\section{Component separation techniques}
\label{sec:cpt_sep}

In the present section we outline the problem of component separation
and briefly describe the techniques used for that purpose in this
work. 
Our aim is to reconstruct the different components of the microwave
sky, and in particular the CMB, in the presence of instrumental noise
from multifrequency microwave observations.

Let us assume that we are performing a multifrequency observation (at
$n_f$ frequencies) of the microwave sky in a given direction $\mathbfss{x}$.
We will obtain a $n_f$-dimensional data vector that contains the
observed temperature fluctuations in this direction at each observing
frequency, plus instrumental noise. The measured data at the $\nu$th
frequency in the direction $\mathbfss{x}$ can be written as
\begin{equation}
d_{\nu}(\mathbfss{x})= \sum_{j=1}^{N_p}B_\nu(|\mathbfss{x}-\mathbfss{x}_j|)
\sum_{p=1}^{n_c} F_{\nu p}\,s_p(\mathbfss{x}_j)\,
+ \eta_\nu(\mathbfss{x}_j)
+ \epsilon_\nu(\mathbfss{x}),
\label{eq:datadef}
\end{equation}
where $n_c$ denotes the number of physical components to be separated
and $N_p$ is the number of pixels in each map.  The function $B_\nu$
accounts for the instrumental beam and $\epsilon_{\nu}(\mathbfss{x})$
corresponds to the instrumental noise at frequency $\nu$ and position
$\mathbfss{x}$.  As is usual for the component separation techniques,
we make the assumption that each of the components (except point
sources) can be factorised into a spatial template ($s_p$) at a
reference frequency $\nu_0$ and a frequency dependence encoded in
$F_{\nu p}$.  Note that point sources can not be factorised in this
way since each of them has its own frequency dependence. The term
$\eta_\nu$ includes the emission of this component (convolved with the
beam) at each frequency.  Therefore, the component separation
technique will attempt, at least, to extract the CMB from the data
$d_{\nu}$ and, possibly, to reconstruct the rest of microwave
components as well as a catalogue of point sources at each frequency.

In the next subsections we outline some of the existing methods to
perform this component separation. We have considered for this work
the maximum-entropy method (MEM), a MEM+MHW joint analysis, the Wiener
filter and a method based on the linear combination of frequency
channels (LCFC).  There are other methods that have been proposed in the
literature. For instance, the EM algorithm of~\cite{mar03}, that
extracts only the CMB, and SMICA~\citep{del03}, which reconstruct all
the components, are blind source approaches that first recover the
power spectra of the considered components and then uses this
information to form the image using Wiener filter.  Therefore they are
also, somewhat, included in our study.  Another important technique
that has been applied to the problem of component separation in the
CMB field is FastICA \citep{mai02}.  Although this method is indeed
very promising, some further development to improve the way in which
FastICA deals with correlated foreground components, realistic beams
and instrumental noise is, however, still needed.  For this reason, we
have chosen not to include this technique in this first analysis of
the effect of component separation on the temperature distribution of
the CMB.

\subsection{The maximum-entropy method}
\label{sec:mem}
The maximum-entropy method (MEM) has been successfully applied to
reconstruct the microwave components from simulated Planck data in a
small patch of the sky \citep{hob98}. The algorithm was extended to
deal with point sources \citep{hob99a} and also combined with the
Mexican hat wavelet (MHW; ~\citealt{vie01b}). Subsequently, MEM has
been adapted to work with spherical data at full Planck resolution
\citep{sto02}. Finally, \cite{sto05} have presented an
improved MEM algorithm that can accommodate 
anisotropic noise and spatial spectral variations. 
All the previous algorithms work in harmonic (or Fourier) space. In
addition, \cite{bar04} developed a MEM 
algorithm that combines both harmonic and real spaces and that can
naturally deal with anisotropic noise and incomplete sky coverage. The
method was applied to the COBE data. \cite{ben03b} has also applied a
different maximum-entropy based algorithm to the WMAP data.

In this subsection we briefly outline the basics of MEM.
For a more detailed derivation, see \cite{hob98,hob99a}.
If the beam is circularly symmetric and assuming that Fourier modes
are independent, it is convenient to work in Fourier space (or
harmonic space if 
dealing with all-sky observations). In this way,
the reconstruction can be performed for each mode separately, what
greatly simplifies the problem. Using matrix notation;
equation (\ref{eq:datadef}) can be written for each Fourier mode as
\begin{equation}
{\mathbfss d}={\mathbfss R}{\mathbfss s} + \boldeta + \bepsilon =
{\mathbfss R}{\mathbfss s} + \bzeta
\label{eq:dataft}
\end{equation}
where $\mathbfss {d}$, $\boldeta$ and $\bepsilon$ are column vectors
each containing $n_f$ complex components and $\mathbfss{s}$ is a
column vector containing $n_c$ complex components.
The response matrix ${\mathbfss R}$ has dimension $n_f \times n_c$
and accounts for the effect of the beam and the spectral dependence of
each of the components. MEM does not attempt to directly reconstruct the point
sources but instead tries to prevent that this emission contaminates
the rest of the components. Therefore, the point source contribution
is just included in the formalism as an extra noise term. This is
reflected in the second equality, where the instrumental noise and
the point source emission has been combined in a general noise
contribution. 

As discussed in \cite{hob98}, MEM should not itself induce
correlations between the elements of the reconstructed vector. 
However, the microwave  components may well be correlated. In
addition, if prior information is available we may wish to include it
in the algorithm. In order to fulfill these requirements, the
reconstruction is performed in terms of a vector
of `hidden' variables $\mathbfss{h}$. The vector of physical 
variables $\mathbfss{s}$ is subsequently found as
\begin{eqnarray}
\label{eq:hidden}
\mathbfss{s} &=& \mathbfss{L} \mathbfss{h} \,\,\,,\\
\label{eq:cholesky}
\mathbfss{C} &=& \mathbfss{L} \mathbfss{L}^T \,\,\,,\\
\mathbfss{C} &=& \left< \mathbfss{s}(k) \mathbfss{s}^\dag(k)\right> \,\,\,
\label{eq:covariance}
\end{eqnarray}
where the dagger denotes the Hermitian conjugate. 
$\mathbfss{C}$ is the signal covariance matrix,  
which contains at the $p$-th diagonal element, the value of the power spectrum
of the $p$-th physical component at the reference frequency $\nu_0$,
whereas the off-diagonal elements contain the cross-power spectrum
between the different components. $\mathbfss{L}$ is a $n_c \times n_c$
lower triangular matrix, obtained by performing a Cholesky
decomposition of $\mathbfss{C}$.
Therefore, we can include in $\mathbfss{L}$  our prior knowledge, if
available, of the power spectra and cross-power spectra
of the microwave components. Note that ${\mathbfss L}$ itself can be
iteratively determined by the MEM \citep{hob98}.
That is, a first reconstruction is obtained using an initial guess for 
${\mathbfss L}$ and subsequently the power spectra of those
reconstructions is estimated. These new power spectra are then used as
a starting point for the next iteration and the process is repeated until
convergence is achieved.

Following Bayes' theorem, we choose as the estimator $\hat{\mathbfss{h}}$
of the hidden vector to be the one that maximises the posterior
probability given by
\begin{equation}
\Pr({\mathbfss h}|{\mathbfss d}) 
\propto \Pr({\mathbfss d}|{\mathbfss h})\Pr({\mathbfss h})
\label{eq:bayes}
\end{equation}
where $\Pr({\mathbfss d}|{\mathbfss h})$ is the likelihood of obtaining
the data given a particular hidden vector and 
$\Pr({\mathbfss h})$ is the prior probability that contains our
expectations about the hidden vector before acquiring any data.

To construct the likelihood function, we assume that the `generalised'
noise contribution (containing instrumental noise and emission from
point sources) is well described by a Gaussian distribution, which has
been shown to work reasonably well \citep{hob99a}. Therefore, the
likelihood function is given by
\begin{eqnarray}
\Pr({\mathbfss d}|{\mathbfss h}) 
& \propto & \exp \left(-\bzeta^\dag {\mathbfss N}^{-1} \bzeta \right)
\nonumber \\
& \propto & \exp \left[-({\mathbfss d}-{\mathbfss RLh})^\dag 
{\mathbfss N}^{-1} ({\mathbfss d}-{\mathbfss RLh})\right] \nonumber \\
& \propto & \exp \left[-\chi^2(\mathbfss h) \right]
\label{eq:likelihood}
\end{eqnarray}
where in the second line we have used (\ref{eq:dataft}). 
The noise covariance matrix ${\mathbfss N}$ has 
$n_f \times n_f$ elements and at each $\bmath{k}$-mode is given
by
\begin{equation}
{\mathbfss N}(\bmath{k}) = \langle\bzeta(\bmath{k})
\bzeta^\dagger (\bmath{k})\rangle.
\end{equation}
Therefore, at any given Fourier mode, the $\nu$th diagonal element of
${\mathbfss N}$ contains the power spectrum at that
mode of the instrumental noise plus the point source contribution to
the $\nu$th frequency channel. The off-diagonal terms account for the
cross-power spectra between different channels. Note that if the noise is
uncorrelated between channels, only the point sources contribute to
the off-diagonal elements.

Regarding $\Pr({\mathbfss h})$, MEM assumes an entropic prior
probability for the hidden vector $\mathbfss{h}$ of the form
\begin{equation}
\Pr({\mathbfss h}) \propto \exp[\alpha S({\mathbfss h},{\mathbfss m})]
\label{eq:prior}
\end{equation}
where $S({\mathbfss h},{\mathbfss m})$ is the cross entropy of the
complex vectors $\mathbfss{h}$ and $\mathbfss{m}$, where ${\mathbfss
m}$ is a model vector to which ${\mathbfss h}$ defaults in absence of
data. $\alpha$ is a regularising parameter that can be fixed in a
fully Bayesian manner. The expression for the cross entropy for
complex images and the method for determining $\alpha$ are discussed
in \cite{hob98}.

Taking into account equations (\ref{eq:likelihood}) and
(\ref{eq:prior}), maximising the posterior probability 
$\Pr({\mathbfss d}|{\mathbfss h})$ with respect to ${\mathbfss h}$ is
equivalent to minimising the function
\begin{equation}
\Phi_{\rm MEM}({\mathbfss h})=\chi^2({\mathbfss h}) 
- \alpha S({\mathbfss h},{\mathbfss m}).
\end{equation}
 Therefore the MEM reconstruction at each Fourier mode is given by the
 vector of hidden variables ${\mathbfss h}$ that minimises the
 previous equation.

\subsection{MEM+MHW joint analysis}

As mentioned in the previous section, MEM does not directly attempt to
reconstruct a catalogue of point sources at each frequency due to the
fact that this emission can not be factorised in a spatial template
and a frequency dependence. Instead, point sources are included as an
extra noise term with the aim of preventing them from contaminating
the reconstructions of the rest of the microwave components.  This
approach has shown to work reasonably well for point sources with low
to intermediate fluxes, greatly reducing the contamination due to this
emission in the reconstructed CMB and the other microwave components
\citep{hob99a}. However, this method does have some limitations. MEM
does not succeed on removing the brightest point sources, which are
present in the reconstructions although with significantly reduced
amplitudes.  This is not surprising, since point source emission is
modelled as an additional Gaussian noise and the brightest point
sources are not well characterised by such a model. In order to solve
this problem, the MEM has been combined with the MHW \citep{vie01b}.

The MHW technique has been developed for the extraction of point
sources from microwave maps. The method was first introduced in
\cite{cay00} and then improved in \cite{vie01a} by including the concept
of optimal scale. The technique was then extended to deal with
spherical data using the spherical Mexican hat wavelet in \cite{vie03},
which provide a predicted catalogue of point sources
from all-sky full-resolution Planck data.

In this section we outline the basics of the MHW technique and how to
combine it with the MEM (for a more detailed derivation see
\citealt{vie01a,vie01b}). The two-dimensional MHW is given by 
\begin{equation}\label{eq:MHW}
\psi(x) = \frac{1}{\sqrt{2\pi}}\Big[2-\big(\frac{x}{R}\big)^2\Big]
        e^{-\frac{x^2}{2R^2}}.
\end{equation}
where $R$ is the wavelet scale.
By convolving the data with the MHW, we obtain the so-called wavelet
coefficients. The main idea behind the MHW technique is that point
sources are amplified in wavelet space and, therefore, they can be
better detected. Wavelet coefficient maps can be obtained for each
wavelet scale $R$. Let us assume that a point source of amplitude $B$
has been convolved with a Gaussian beam of dispersion $\sigma_a$. The
value of the wavelet coefficient for scale $R$ at the position of the
source is given by
\begin{equation}\label{eq:Coeff}
   \frac{w(R)}{R} = 2\sqrt{2\pi}\frac{B}{A}
                              \frac{(R/\sigma_a)^2}
                              {(1 + (R/\sigma_a)^2)^2},
\end{equation}
where $A$ is the area occupied by the point source.

The variance of the wavelet coefficient map at scale $R$ can be
calculated as
\begin{equation}\label{eq:SW}
   \sigma_{w}^2(R) = 2\pi R^{2}\int dkP(k)k |\widehat{\psi}(Rk)|^2,
\end{equation}
where $P(k)$ is the power spectrum of the map to be analysed and
$\widehat{\psi}$ corresponds to the Fourier transform 
of the MHW.

As already mentioned, point
sources are amplified in wavelet space, i.e., the level of the point
sources relative to the dispersion of the map is higher in wavelet space
than in real space:
\begin{equation}\label{eq:Ampl}
   \frac{w(R)}{\sigma_{w}(R)} > \frac{(B/A)}{\sigma},
\end{equation} 
where $\sigma$ is the dispersion of the observed map (dispersion in
real space).  In order to obtain the best conditions to detect point
sources, the first term in the previous equation should be as high as
possible. To do this, the method looks for an optimal scale $R_{\rm
opt}$ that maximises that quantity and therefore gives the maximum
amplification of the point sources in wavelet space. Taking into
account equations (\ref{eq:Coeff}) and (\ref{eq:SW}) $R_{\rm opt}$ can
be easily obtained directly from the data, since the optimal scale
only depends on $\sigma_a$ and the power spectrum of the map to be
analysed. As expected, $R_{\rm opt}$ takes values around $\sigma_a$,
since this is the scale that characterises the point sources. The
background of the image also plays a role in determining $R_{\rm
opt}$. Backgrounds with high power at scales larger than the beam
size, will tend to move $R_{\rm opt}$ towards smaller scales than
$\sigma_a$ and vice-versa.

The procedure to detect point sources is as follows.
First, the optimal scale is found and the map is filtered with the MHW
of that scale. All pixels above a $5\sigma_w(R_{\rm opt})$ threshold
are considered  
as point sources candidates. The data map is additionally filtered with
another three adjacent scales. Those four wavelet coefficients maps are
used to estimate the 
amplitude of the source by fitting
the `experimental' $w(R)$ curve at the position of the candidate to the
expected theoretical values given by equation (\ref{eq:Coeff}).
If the fit is good, the candidates are
accepted as point sources, if not, they are discarded.

The combination of the MEM and MHW consists on applying sequentially
each of the methods. First of all, the MHW is applied to each of the
frequency maps, which allows one to detect the brightest point
sources. These sources are subtracted from the original data and
MEM is then applied to these `cleaned' maps, modelling the
unsubtracted point sources as an additional noise term.
Since the MHW can successfully remove the brightest sources and MEM can
deal with those with low to intermediate flux, both methods
complement each other and the reconstructions of the different
microwave components are significantly improved when combined.
Moreover, an improved point sources catalogue at each frequency map
can also be obtained with the following procedure.
The reconstructed maps are used to generate `mock' data which
are subtracted from the original ones. This provides
residual maps at each frequency channel which contain mainly
contributions from point sources and instrumental noise plus some
residuals from other 
emissions. The MHW is then applied to these residual maps to produce
a point source catalogue at each frequency that reaches lower fluxes
and errors than those obtained using only the MHW.

\subsection{The Wiener filter}

Wiener filter is defined as the linear filter that minimises the
variance of the errors of the reconstruction (e.g. \citealt{ryb92}).
This technique was generalised to deal with multifrequency and
multiresolution microwave data in \cite{bou96} and \cite{teg96}.
\cite{hob98} derived the Wiener filter in a
Bayesian context and compare its performance with MEM when
reconstructing the microwave sky from simulated Planck
observations, showing that MEM outperformed WF.

Following \cite{hob98} we will derive the Wiener filter
within the Bayesian framework. In this context,
the solution of the Wiener filter is found by assuming that the
probability distribution of the sky emission is well described by a
multivariate Gaussian characterised by a given covariance
matrix. Therefore, the probability distribution of the vector
$\mathbfss{s}$ at each Fourier mode is also described by a
multivariate Gaussian distribution of dimension $n_c$. This leads to a
prior probability of the form 
\begin{equation}
\Pr({\mathbfss h}) \propto \exp\left(-{\mathbfss s}^\dag{\mathbfss
C}{\mathbfss s}\right) 
\end{equation}
where ${\mathbfss C}$ is the signal covariance matrix defined in (\ref{eq:covariance}).
Taking into account the form of the likelihood found previously
(equation \ref{eq:likelihood}), the posterior probability is given by
\begin{equation}
\Pr({\mathbfss s}|{\mathbfss d})
\propto \exp\left[-\chi^2({\mathbfss s}) 
-{\mathbfss s}^\dag{\mathbfss C}{\mathbfss s}\right]
\end{equation}
or, equivalently,
\begin{equation}
\Pr({\mathbfss s}|{\mathbfss d})
\propto \exp\left[-\chi^2({\mathbfss h}) 
-{\mathbfss h}^\dag{\mathbfss h}\right]
\end{equation}
where we have made used of equations (\ref{eq:hidden}-\ref{eq:cholesky}).
Therefore, the Wiener reconstruction at each Fourier mode can be found
by minimising, with respect to the hidden vector ${\mathbfss h}$,
the function
\begin{equation}
\Phi_{\rm WF}({\mathbfss h})=\chi^2({\mathbfss h}) 
+{\mathbfss h}^\dag{\mathbfss h}
\end{equation}
and then obtain the physical components as ${\mathbfss s}= {\mathbfss L}
{\mathbfss h}$.

We would like to point out that \cite{hob98} showed that WF can be
seen as an approximation to MEM in 
the small fluctuation limit and, therefore, for Gaussian signals, both
reconstructions become very similar.

\subsection{Linear combination of frequency channels}

If we assume that the foreground contamination is small in comparison
with the CMB signal, a simple approach that combines linearly the different
frequency channels (LCFC) in order to increase the signal-to-noise ratio can
be used. In this case, the estimation of the cosmological signal in the
sky at position $\mathbfss{x}$ is given by
\begin{equation}
\hat{s}_{\rm
CMB}(\mathbfss{x})=\sum_{j=1}^{N_m}w_j(\mathbfss{x})d_j(\mathbfss{x}) 
\end{equation}
where $N_m$ is the number of frequency maps to be combined (we do not
necessarily include all the available frequency channels) and the data
$d_j$ are 
given in thermodynamic temperature. In order to reduce the effect of
instrumental noise, the weights of the map are chosen
to be inversely proportional to the noise variance at each position of
the sky and therefore the coefficients $w_j(\mathbfss{x})$ are given by:
\begin{equation}
w_j(\mathbfss{x})=\frac{1}{\sigma_j^2(\mathbfss{x})}
\left[\sum{\frac{1}{\sigma_j^2(\mathbfss{x})}}\right]^{-1}
\end{equation}
where $\sigma_j^2(\mathbfss{x})$ is the noise dispersion of the
frequency map $j$ at position $\mathbfss{x}$.  Note that this
procedure provides a map with increased signal-to-noise ratio and
well-known noise properties but it does not attempt to remove any
foreground contamination. However, when dealing with real data, some
subtraction of foregrounds will also be performed in the combined
map. For instance the WMAP team has used this type of combination to
produce a map on which performing Gaussianity studies
\citep{ben03b,kom03} but, before combining the maps, a fit to Galactic
templates was removed from each of them in order to remove foreground
contamination . In particular, they use all the receivers at Q, V and
W bands (at 41, 61 and 94 GHz), since they are dominated, outside the
Galactic plane, by the cosmological signal. On the contrary, they avoid to
include in the combination the K and Ka bands (at 23 and 33 GHz) which
contain significant Galactic emission.

For the sake of simplicity, we will mimic the procedure of template
fitting by reducing the Galactic foregrounds present in each frequency
channel, prior to the linear combination, down to a 10 per cent of its
total amplitude. Therefore, when referring to the LCFC method, unless it is 
otherwise stated, we will assume that this foreground removal step
has been previously performed.

\section{The Gaussianity test}
\label{sec:wavelets}
Given the importance of analysing the temperature distribution of the
CMB fluctuations, many methods have been developed to perform
Gaussianity studies. These include, among others, the Minkowski functionals
(\citealt{col98,got90,kog96,kom03}), the bispectrum
(\citealt*{fer98,hea98,mag00}), properties of hot and cold spots
(\citealt{col87,mar00}), geometrical estimators
(\citealt{bar01,dor03,mon05}), extrema 
correlation function (\citealt{nas95,kog96,bar98,hea99})
goodness of fit tests \citep{cay03,ali03,ali05},
multifractals \citep{pom95}, partition function 
(\citealt{die99, mar00}), phase analysis
(\citealt{chi03};~\citealt*{chib04};~\citealt{col04}) and higher
criticism statistic \citep{cay05}.

In addition, wavelet techniques have been introduced in the last years
for CMB analyses for both small patches of the sky and spherical data.
\cite*{fer97} investigate a set of statistics based on cumulants and
defined in wavelet space. \cite*{hob99} study the power of the
cumulants of the distribution of wavelet coefficients at each scale to
detect cosmic strings on small patches of the sky. This work was
extended by \cite{barh01} who compare the performance of different
algorithms to construct 2-dimensional wavelets. \cite{for99} proposed
a similar method that was tested on simulated CMB maps containing
secondary anisotropies \citep{agh99} and on the COBE-DMR data
\citep*{agh01}. The skewness, kurtosis and scale-scale correlation
coefficients of the COBE-DMR data in the QuadCube pixelization using
planar orthogonal wavelets was carried out by \cite*{pan98}. This
analysis was extended by \cite*{muk00}.  Recently, wavelet methods
have been adapted to deal with spherical data. \cite{bar00} and
\cite{cay01} studied the skewness, kurtosis and scale-scale
correlation of the COBE-DMR data in HEALPix pixelization using the
Spherical Haar Wavelet and the Spherical Mexican Hat Wavelet (SMHW),
respectively. \cite{mar02} compared the performance of these two
spherical wavelets to detect non-Gaussianity. \cite{cay03b} put
constraints on the $f_{NL}$ parameter from the COBE-DMR data using the
SMHW. An analysis based also on the SMHW has been performed on the
WMAP data by \cite{vie04} finding a non-Gaussian signature at scales
of $\sim 10^\circ$.  This work has been extended by \cite{muk04},
\cite{cru05} and \cite{mce05}. Finally, steerable wavelets on
the sphere have been recently proposed by \cite{wia05a,wia05b} to perform
Gaussianity analysis of the CMB. A comparison of the
performance of wavelet methods with other techniques can also be seen
in \cite{agh03} and \cite{cab04b}.

In the present work, we will use a wavelet-based method to detect
non-Gaussianity following the ideas of \cite{hob99} and \cite{barh01}.
In this section we outline the basics of the method. For a more
detailed description of the method, see the previous works. 
The wavelet transform has been extensively described elsewhere
(e.g. \citealt{dau92}). For instance, \cite*{bur98} 
give a detailed introduction to wavelets. The basic idea behind the
wavelet transform is to decompose the considered signal in a series of
wavelet coefficients that keep simultaneous information of real and
Fourier space. Therefore each coefficient can be associated to a
position $l$ and scale $j$ of the image, what makes this analysis very
useful in many applications. In general the position and scale
parameters can take continuous values. However, for pixelized images
is more convenient to restrict these parameters to take a set of
discrete values. In particular, one can construct a discrete set
of wavelets that act as a complete basis for the digitised image.

In 1-dimension, the wavelet basis is constructed from dilations and
translations of the mother (or analysing) wavelet function $\psi$ and a
second related function called the father (or scaling) wavelet
function $\phi$: 
\begin{eqnarray}
\psi_{j,l} & = & 2^{\frac{j-J}{2}}\psi\left( 2^{j-J}x-l\right) \, , 
\nonumber \\
\phi_{j,l} & = & 2^{\frac{j-J}{2}}\phi\left( 2^{j-J}x-l\right) \, , 
\end{eqnarray}
where $2^J$ is the number of pixels of the considered discrete signal $s(x_i)$
and $ 0 \ge j \ge J-1$ and $ 0 \ge l \ge 2^J-1 $ denote
the dilation and translation indices, respectively.

There is not a unique choice for the wavelet basis. For this work, we
will use the so-called Daubechies-4 wavelets that form a real, orthogonal
and compactly supported wavelet basis (see \citealt{dau92} for the
derivation of these functions).
The discrete signal $s$ at pixel $x_i$ can then be written
\begin{equation}
s(x_i)= a_{0,0}\phi_{0,0}(x_i) + \sum_j \sum_l d_{j,l}\psi_{j,l}(x_i)
\, ,
\end{equation}
$a$, $d$ correspond to the approximation and detail wavelet coefficients
respectively. These coefficients can be obtained in a recursive way
starting from the data vector $f(x_l) \equiv a_{J,l}, l=0,..,2^{J}-1$
\begin{eqnarray}
a_{j-1,l} & = & \sum_m h(m-2l)a_{j,m} \nonumber \\
d_{j-1,l} & = & \sum_m g(m-2l)a_{j,m} \, .
\end{eqnarray}
where $h$, $g$ are the low and high-pass filters associated to the scaling
and analysing wavelet functions through the refinement equation (see
e.g. \citealt{bur98}). 

At each iteration, the vector of
length $2^j$ is split into two components: $2^{j-1}$ detail coefficients
and the same number of approximation coefficients. These 
approximation coefficients are then used 
as input for the next iteration to construct the detail and
approximation components at the next larger scale.
The process is repeated down to the lowest resolution level considered.
This leads to a number of wavelet
coefficients equal to the original number of pixels.
As the index $j$ decreases from $J-1$ to 0, the wavelet
coefficients give information about the structure of the function on 
increasingly larger scales, with each scale a factor of 2 larger than
the previous one. 

The extension of the discrete wavelet transform to two-dimensions is
not unique. Following \cite{mal89} we will construct two-dimensional
bases as tensor products of the scaling and/or analysing wavelet at
each scale $j$:
\begin{eqnarray*}
\phi_{j;l_1,l_2}(x,y) & = & \phi_{j,l_1}(x)\phi_{j,l_2}(y) \\
\psi^{\rm H}_{j;l_1,l_2}(x,y) & = & \psi_{j,l_1}(x)\phi_{j,l_2}(y) \\
\psi^{\rm V}_{j;l_1,l_2}(x,y) & = & \phi_{j,l_1}(x)\psi_{j,l_2}(y) \\
\psi^{\rm D}_{j;l_1,l_2}(x,y) & = & \psi_{j,l_1}(x)\psi_{j,l_2}(y).
\end{eqnarray*}
where H, V and D stand for horizontal, vertical and diagonal.
The wavelet basis constructed in this way is non-redundant and
orthogonal.

Let us assume that we have an image with $2^J \times 2^J$ pixels.
With this scheme, the image (scale $J$) is decomposed into an
approximation and three detail (horizontal, vertical and diagonal)
images corresponding to scale $J-1$, with each of the images
containing $2^{J-1} \times 2^{J-1}$. Note that the approximation image
is basically a smoothed version of the input map, whereas the detail
coefficients keep the information of the difference between the
original and smoothed images. The algorithm is then applied again to
the approximation coefficients at scale $J-1$ to produce the
approximation and detail coefficients at scale $J-2$ and so on.  Note
that the coefficients at scale $j$ keep information of the structure
of the image at scales approximately equal to the pixel size times
$\sim 2^{J-j}$.

In order to test the Gaussianity
of the CMB, the considered temperature map is wavelet transformed
using Daubechies-4. A certain statistic is then calculated at each
wavelet scale 
for each of the three types of wavelet coefficients.
In particular, following \cite{hob99}, we will consider the
fourth order cumulant, that can be estimated as
\begin{eqnarray}
\hat{\kappa}_4(j,T) & = &
\frac{N_j^2[(N_j+1)\hat{\mu}_4-3(N_j-1)\hat{\mu}_2^2]} 
{(N_j-1)(N_j-2)(N_j-3)} \nonumber \\
\hat{\mu}_r(j,T) &=& \langle (d_j^T-\langle d_j^T \rangle )^r \rangle.
\end{eqnarray}
where $\hat{\kappa}_4(j,T)$ corresponds to the fourth cumulant estimated
at scale $j$ for the $T$ type of detail coefficient,
$d_j^T$ are the detail wavelet coefficients, 
$N_j$ is the number of considered coefficients at scale $j$ and
$\hat{\mu}_r(j,T)$ are the estimated central moments.
The same procedure is repeated for a large number of Gaussian
realisations (5000) with the same power spectrum as the test map.
The $\hat{\kappa_4}$ value obtained for the input map is then compared with 
the distribution of $\hat{\kappa_4}$ obtained from the Gaussian realisations.
Departures from this distribution will indicate a detection of
non-Gaussianity. 
To avoid spurious boundary effects we do not include in the analysis 
those coefficients that contain information from pixels close to the
borders (in particular, we do not consider ten rows/columns in each
border).

\section{The simulated data}
\label{sec:simulations}
In order to test the effect of component separation on the
distribution of the CMB, we have generated simulated observations of
small patches (12.8$^\circ \times$ 12.8$^\circ$) of the microwave sky
according to the characteristics of the Planck satellite (see
table~\ref{tab:planck}).  
\begin{table}
   \begin{center}
         \begin{tabular}{|c|c|c|c|c|}
	 \hline
	 Frequency & Fractional & FWHM & Pixel size & $\sigma_{noise}$\\
	 (GHz) & bandwidth & (arcmin) & (arcmin) & $(\times 10^{-6})$ \\
\hline
	 30 & 0.2 & 33.0 & 6.0 & 2.0 \\
	 44 & 0.2 & 24.0 & 6.0 & 2.7 \\
	 70 & 0.2 & 14.0 & 3.0 & 4.7 \\
	 100 & 0.33 & 9.2 & 3.0 & 2.0 \\
	 143 & 0.33 & 7.1 & 1.5 & 2.2 \\
	 217 & 0.33 & 5.0 & 1.5 & 4.8 \\
	 353 & 0.33 & 5.0 & 1.5 & 14.7 \\
	 545 & 0.33 & 5.0 & 1.5 & 147 \\
	 857 & 0.33 & 5.0 & 1.5 & 6700 \\
	 \hline
      \end{tabular}
      \caption{\label{tab:planck}Characteristics of the Planck
      satellite at the 9 
      frequency channels. The central frequency and fractional
      bandwidth are given in columns 1 and 2. The
      antenna full-width half-maximum (FWHM)  is given in column 3
      (a Gaussian pattern is assumed for all channels). The
      characteristic pixel sizes are shown in column 4. Finally, the
      fifth column gives the expected instrumental noise (assumed to
      be Gaussian) in $\Delta T/T$ per resolution element (which is
      assumed to be a square pixel of size the corresponding FWHM).}
    \end{center}
\end{table}
The simulations contain, in addition to the cosmological signal,
Galactic foregrounds (thermal dust, free-free and synchrotron),
thermal and kinematic Sunyaev-Zeldovich effects, extragalactic point
sources and Gaussian white noise.

\begin{figure*}
\includegraphics[angle=0,width=\hsize]{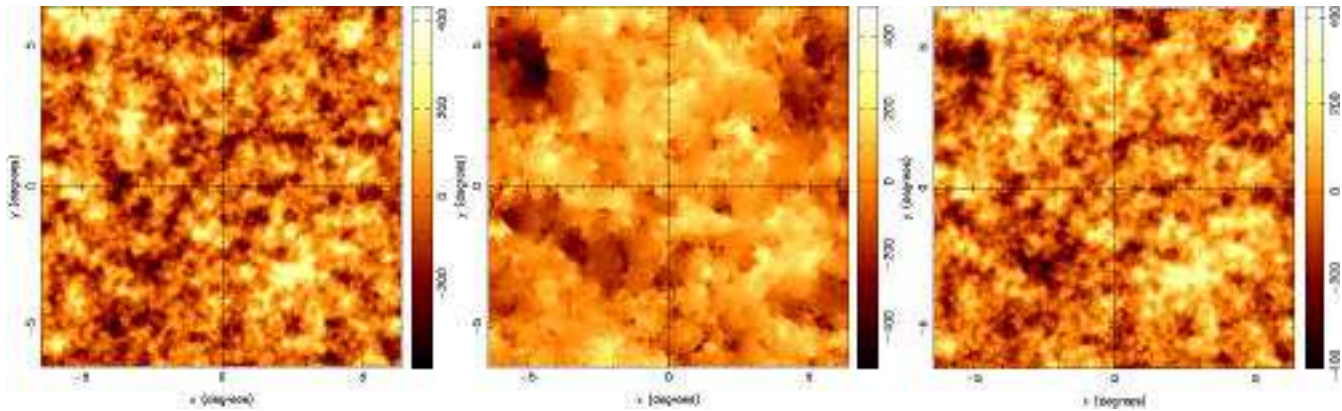}
\caption{Test CMB maps used in the simulations. The different panels
correspond to: Gaussian CMB (left), cosmic strings (middle),
and Gaussian CMB plus cosmic strings in proportion 2:1 (right). 
The maps have been smoothed with a Gaussian beam of fwhm=5$'$ and the
units are thermodynamic temperature in $\mu$K.}
\label{fig:input_cmb}
\end{figure*}

We have used two reference CMB maps for our simulations. On the one
hand, we have generated a Gaussian CMB map assuming the standard
$\Lambda$CDM model (whose power spectrum was generated using CMBFast,
\citealt{sel96}). On the other hand, since we would like to test the
effect of component separation on intrinsic non-Gaussianity, we have
also used a simulated CMB map of cosmic strings \citep*{bou88}. Given
that recent CMB experiments have ruled out pure topological defect
based scenarios (e.g. \citealt{bou01}) we will construct CMB maps with
different proportions of the Gaussian and cosmic strings maps.
Fig.~\ref{fig:input_cmb} shows the Gaussian CMB, the cosmic strings
simulation and a map with a mixture of the two components.

\begin{figure*}
\includegraphics[angle=0,width=\hsize]{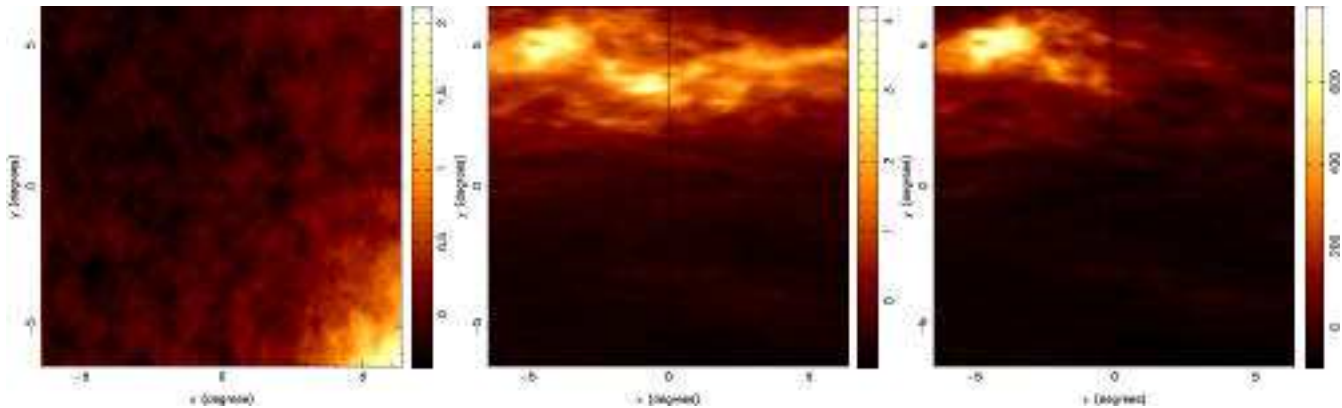}
\caption{Simulated synchrotron (left), free-free (middle) and dust
(right) emissions for our reference zone (region 3) at 300 GHz. The
maps have been smoothed with a Gaussian beam of fwhm=5$'$ and the
units are thermodynamic temperature in $\mu$K.}
\label{fig:foregrounds}
\end{figure*}

As Galactic foregrounds we have included simulations of synchrotron,
thermal dust and free-free emissions, which have been simulated at the
frequency of 353 GHz. The different emissions have been simulated
assuming that the frequency dependence is spatially constant. The
synchrotron template at this frequency has been obtained using the
model of \cite{gia02}.  To simulate this component at the different
Planck frequencies we have rescaled this template using a power law
$I_\nu \propto \nu^{-\alpha_{\rm syn}}$ with $\alpha_{\rm syn}=-0.9$.
The thermal dust template has been generated using the model of
\cite*{fin99}. For simplicity, we have then assumed a simple grey body
law with parameters $T_d=18 K$ and $\beta_{\rm emis}=2.0$ to simulate
the contribution of this component at the Planck frequency
maps. Finally, the free-free has been simulated using the correlation
with dust emission in the way proposed by \cite{bou99}. The template
has then been extrapolated to the considered frequencies using a power
law with $\alpha_{\rm ff}=-0.16$. In order to test the robustness of
the results we have tested the methods using four different Galactic
regions of the sky with $|b| > 20^\circ$.
The dispersion of the three Galactic
components at the frequency of 300 GHz for the four considered
regions is given in Table~\ref{tab:regions}. 
Two of the regions (1 and 3) are within the brightest 2 per cent of
the sky in dust emission, whereas the other two regions (2 and 4)
correspond to more typical regions with lower Galactic dust emission.
Fig.~\ref{fig:foregrounds} shows the
synchrotron, free-free and dust components for one of the considered
regions of the sky.
\begin{table}
   \begin{center}
         \begin{tabular}{|c|c|c|c|c|}
	 \hline

	 & Region 1 & Region 2 & Region 3 & Region 4 \\
\hline
Dust & 292.9 & 50.33 & 115.8 & 26.7 \\
FF & 1.46 & 0.89 & 0.91 & 0.42 \\
Synch. & 1.29 & 0.09 & 0.28 & 0.70 \\
	 \hline
      \end{tabular}
      \caption{\label{tab:regions} Dispersion of the three Galactic
components for the four considered regions at 300 GHz smoothed with a
Gaussian beam of 5 arcminutes. The units are thermodinamycal
temperature in $\mu$K.}
    \end{center}
\end{table}

The kinematic and thermal SZ effects have been simulated using the
model of \cite{die01} for the standard $\Lambda$CDM model.
The radio and infrared point sources have been simulated
according to the model of \cite{tof98} (for recent improvements of this
model, see \citealt{dez05} and \citealt*{gon05}) for the same cosmological
model . Finally, we have simulated the instrumental noise as
Gaussian white noise according to the Planck characteristics. 

\begin{figure*}
\includegraphics[angle=-90,width=\hsize]{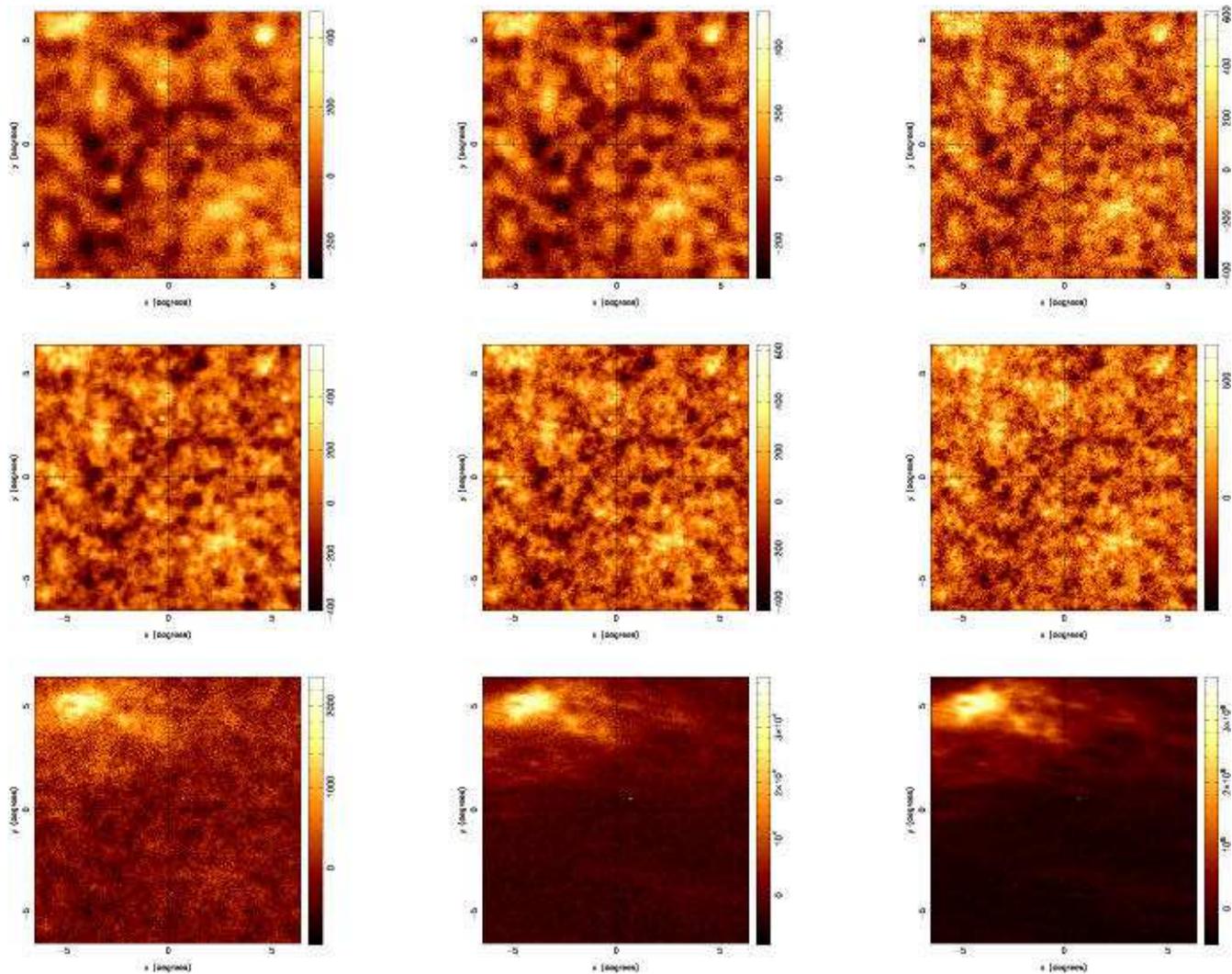}
\caption{Simulated Planck data containing Gaussian CMB, Galactic
synchrotron, dust and free-free emissions, thermal and kinematic
Sunyaev-Zeldovich effects, extragalactic point source emission and
white noise. From left to right and top to bottom the different panels
correspond to 30, 44, 70, 100, 143, 217, 353, 545 and 857 GHz channels.
The units are thermodynamic temperature in $\mu$K.}
\label{fig:data}
\end{figure*}

Planck simulated data generated from the reference Gaussian CMB and
the Galactic foregrounds of Fig.~\ref{fig:foregrounds} (region 3) are
given in Fig.~\ref{fig:data}. 
Table~\ref{tab:rms_cpt} gives the
contribution in rms of each microwave component at each frequency
channel for the same case.

\begin{table*}
   \begin{center}
         \begin{tabular}{|c|c|c|c|c|c|c|c|}
	 \hline
	Frequency (GHz) & CMB & kSZ & tSZ & Dust & Free-free & Synchrotron &
Point sources \\
\hline
30  & 72.8 & 0.29 & 2.15 & 0.25 & 18.6 & 29.6 & 19.9 \\
44  & 83.3 & 0.37 & 2.70 & 0.54 & 8.48 & 10.3 & 13.1 \\
70  & 97.7 & 0.53 & 3.55 & 1.45 & 3.40 & 2.98 & 9.31 \\
100 & 105.7 & 0.67 & 3.78 & 3.35 & 1.80 & 1.23 & 7.75 \\
143 & 109.8 & 0.79 & 2.95 & 8.35 & 1.08 & 0.57 & 5.38 \\
217 & 113.4 & 0.95 & 0.00 & 31.5 & 0.80 & 0.31 & 5.10 \\
353 & 113.4 & 0.95 & 7.17 & 284.8 & 1.18 & 0.32 & 17.1 \\
545 & 113.4 & 0.95 & 17.4 & 5455.5 & 5.10 & 1.00 & 287.7 \\
857 & 113.4 & 0.95 & 33.2 & 525918 & 130.4 & 18.4 & 13588 \\
	 \hline
      \end{tabular}
      \caption{\label{tab:rms_cpt} Dispersion of the different microwave
components in the Planck simulated data maps given in
Fig.~\ref{fig:data} (corresponding to region 3). The units are
thermodinamycal temperature in $\mu$K.}
    \end{center}
\end{table*}

\section{Results}
\label{sec:results}
We have performed our wavelet analysis on the recovered CMB maps
obtained with the three considered reconstruction methods. We have
also studied different input CMB maps.
First of all, we have produced Planck simulated data
using our reference Gaussian CMB map and have studied if the
reconstruction techniques preserve the Gaussian character of the temperature
fluctuations (\S\ref{sec:gaussian}). Secondly, we have considered two
different non-Gaussian 
CMB maps (containing different proportions of cosmic strings) and have
studied if the underlying non-Gaussianity is still detected in the
reconstructed temperature map (\S\ref{sec:ng}). To understand better
the effect of 
component separation techniques on the CMB we have neglected the
effect of point sources in these two first cases, since this
contaminant needs to be treated in a very different way. In a third
case, we include the emission 
coming from point sources in the Planck simulated data and study the
effect of this contaminant when the underlying CMB map is
Gaussian. The study is done on reconstructed maps obtained using MEM
and the MEM+MHW joint analysis (\S\ref{sec:point_sources}).

In order to test the robustness of the results we have obtained the
CMB reconstruction using simulated data of four different Galactic
regions of the sky (see Table~\ref{tab:regions}).  We have found that
the output of the Gaussian analysis is quite insensitive to the choice
of the Galactic region and therefore we present our results only for
simulations constructed using the Galactic templates of
Fig.~\ref{fig:foregrounds}.  We have also tested our results with
different noise realisations, finding that the conclusions of the
analysis are not modified.

For the LCFC, different combinations of frequency maps have been
tried. In particular, we have found that the combination of the 143
and 217 GHz channels is the best choice for the considered cases in
the sense of retaining the underlying CMB temperature
distribution. Therefore, unless it is otherwise stated, we show our
results for this particular combination.

\subsection{Gaussian case}
\label{sec:gaussian}
\begin{figure*}
\includegraphics[angle=0,width=\hsize]{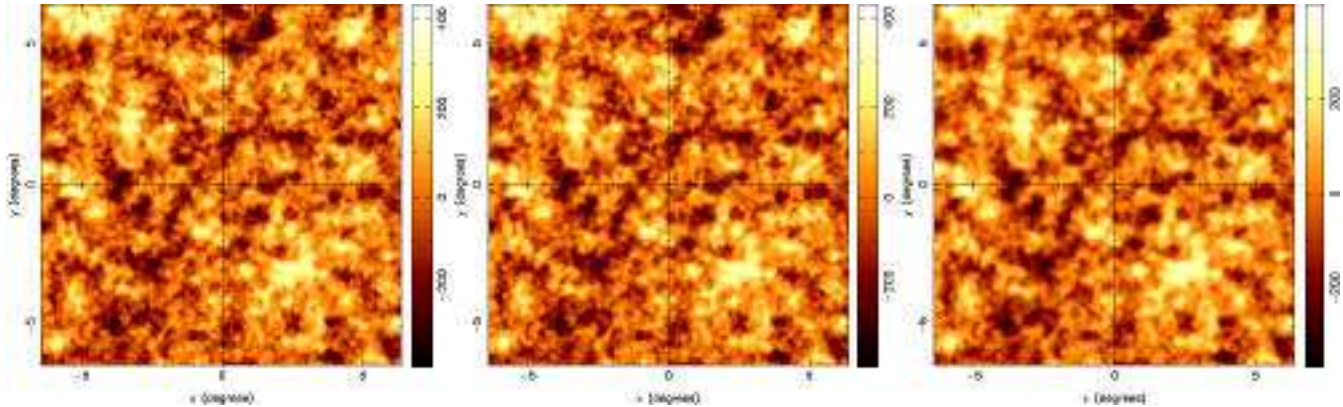}
\caption{Reconstructed CMB map obtained with MEM (left), WF (middle)
and a linear combination of the 143 and 217 GHz channels (right) from
Planck simulated data using the Gaussian CMB as input. The
maps have been smoothed with a Gaussian beam of fwhm=5 arcmin and the
units are thermodynamic temperature in $\mu$K.} 
\label{fig:rec_gaus_cmb_z3}
\end{figure*}

The CMB reconstructed maps obtained from the Planck simulated data for
the input Gaussian CMB are given in
Fig.~\ref{fig:rec_gaus_cmb_z3}. The different panels correspond to the
reconstructions obtained using MEM (left), WF (middle) and the
LCFC using the 143 and 217 GHz
channels (right). These maps should be compared with the input
Gaussian CMB of Fig.~\ref{fig:input_cmb}. Note that the MEM and WF
reconstructed maps look very similar. The LCFC has
a lower resolution than the MEM and WF reconstructions, since
these two methods perform a (partial) deconvolution of the signal.

\begin{figure}
\includegraphics[angle=270,width=\hsize]{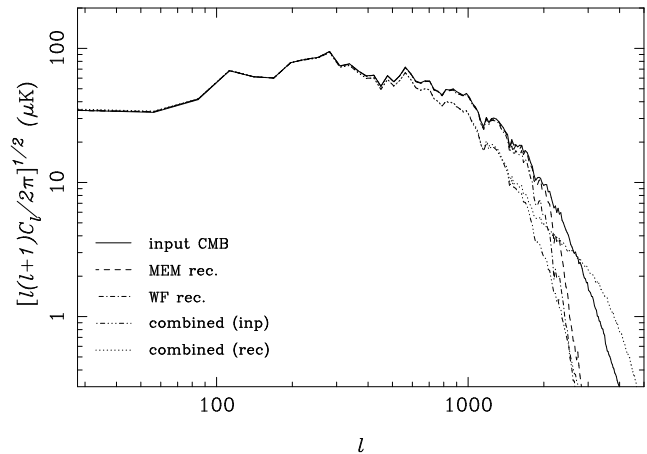}
\caption{Power spectra obtained from the input and reconstructed CMB
  maps for the Gaussian case smoothed with a Gaussian beam of fwhm=5
  arcmin (see text for details).}
\label{fig:ps_gaus_cmb_z3}
\end{figure}

Fig.~\ref{fig:ps_gaus_cmb_z3} shows the power spectra of the previous
reconstructed CMB maps versus the input one. The solid line shows the
power spectrum of the input Gaussian map, given in the left panel of
Fig.~\ref{fig:input_cmb}. As shown in previous works, MEM (dash
line) can recover reasonably well the CMB power spectrum up to $\ell \la$
2000, whereas WF (dot-dash line) starts to underestimate the 
$C_{\ell}$'s around $\ell \sim 1500$. The dot line corresponds to
the LCFC, where we appreciate two differences with respect to the
input power spectrum: a defect of power at intermediate $\ell$'s, since
the combined map has a lower resolution than the one of the input, and an
excess at the higher multipoles, due to the presence of instrumental
noise in the combined map. This curve can also be compared with the
dash-three dots line, which corresponds to the power spectrum of a
combined map obtained from data channels where only CMB is
present. This represents the best reconstruction that could be
achieved with the LCFC. We see that the reconstructed LCFC follows
reasonably well this ideal LCFC power spectrum up to $\ell \sim
1500$. 

\begin{figure}
\includegraphics[angle=0,width=\hsize]{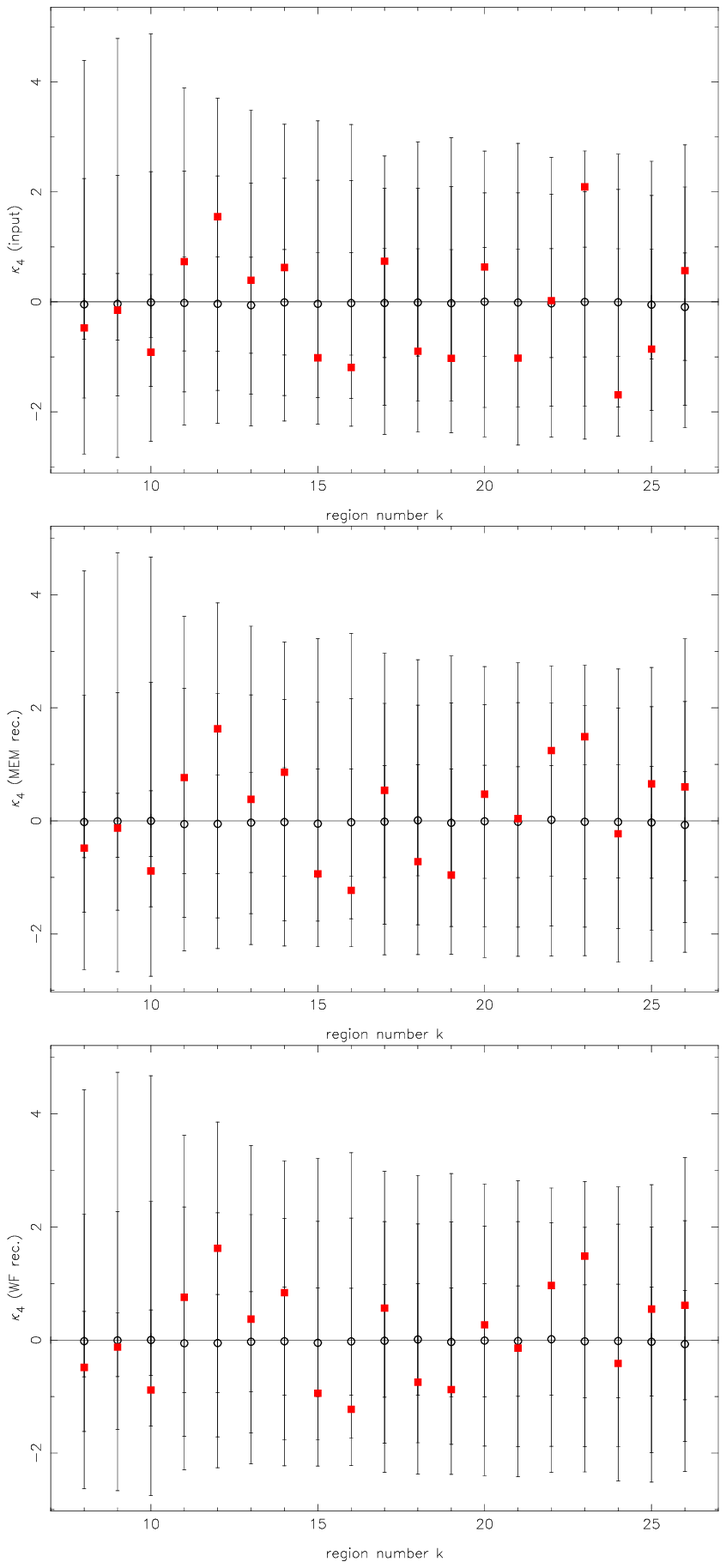}
\caption{Value of $\kappa_4$ versus region number $k$ for the case of
Gaussian CMB. The different panels correspond to the input CMB (top),
MEM reconstruction (middle) and WF reconstruction (bottom). The open
circles and the error bars correspond to the average and the 68, 95
and 99 per cent acceptance regions obtained from 5000 Gaussian
simulations with the same power spectrum as the test map. The solid
squares are the values obtained for the test map.
Note that the small differences in the error bars (of the order of about a
few percent) between the input and reconstructions are due to the
limited number of simulations. In addition, since the reconstruction is not
perfect, the power spectra of the reconstructed maps differ slightly
from the input one, what also introduces small differences in the
distribution of the $\kappa_4$ values.}
\label{fig:k4_gaus_cmb_z3}
\end{figure}

\begin{figure}
\includegraphics[angle=0,width=\hsize]{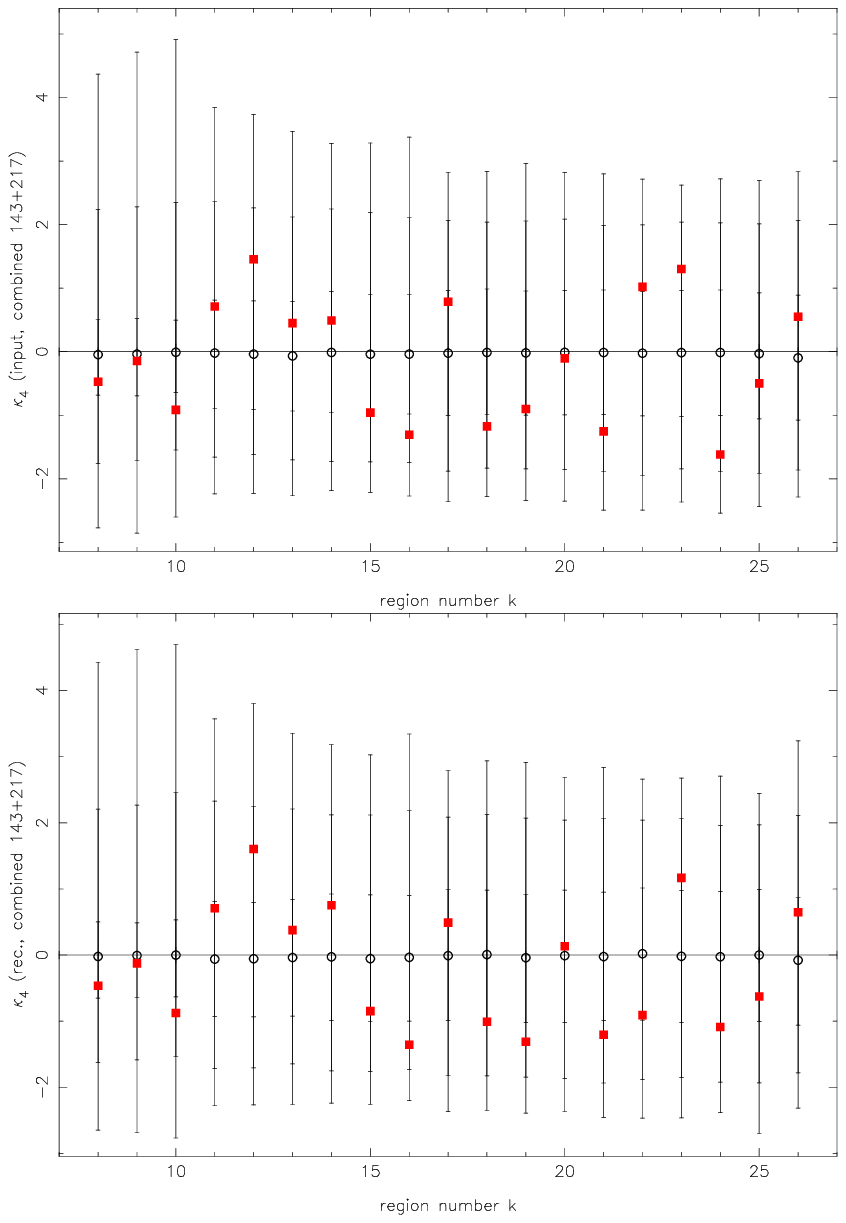}
\caption{Value of $\kappa_4$ versus region number $k$ for the case of
Gaussian CMB. The panels correspond to the input (top) and
reconstructed (bottom) combined map using the 143 and 217 GHz channels}
\label{fig:k4_gaus_cmb_z3b}
\end{figure}

In order to test the effect of the component separation technique
on the underlying distribution of the CMB, we have applied our wavelet
method to these three reconstructed CMB maps and compared the results
with the input one. Figs.~\ref{fig:k4_gaus_cmb_z3}
and~\ref{fig:k4_gaus_cmb_z3b} show the value of $\kappa_4$ (solid
squares) at each wavelet scale for the input and reconstructed CMB maps. 
As already mentioned there are three different kinds of wavelet
coefficients for the type of wavelet that we are using: vertical,
horizontal and diagonal. Horizontal and vertical details should be
statistically equivalent, but we obtain different levels of detection
because we are looking at a particular realisation. The $x$-axis on
the plots gives the numbering $k$ of the regions which goes as
follows: region 1 corresponds to the approximation coefficients at the
lowest resolution; regions 2,3,4 correspond to vertical, horizontal
and diagonal details respectively for the largest scale ($j=1$);
regions 5,6,7 give the three details in the same order for the next
scale ($j=2$) and so on.  The last scale ($j=26$) is not obtained from
wavelet coefficients but corresponds to the value of $\kappa_4$
obtained directly from the temperature
map. Table~\ref{tab:regions_mallat} gives the correspondance between
the region number $k$ with the scale $j$ and type of coefficients. It
also gives an estimation of the scale corresponding to each wavelet
region. Since no coefficients are retained at the lowest regions, the
x-axis runs only from $k=8$ to 26. The open circle and error bars
shown in the figures correspond to the average and the 68, 95 and 99
per cent acceptance regions obtained from 5000 Gaussian simulations
with the same power spectrum as the test map. For comparison purposes
the distributions of $\kappa_4$ at each scale have been normalised to
unit dispersion. Therefore the $y$-axis gives directly the values of
$\kappa_4$ in numbers of $\sigma$ (note, however, that these
distributions are not necessarily Gaussian and thus the number of
$\sigma$'s in general will not correspond to the usual confidence
intervals of the Gaussian case).
\begin{table}
\caption[]{Correspondence of region number $k$ with wavelet scale
$j$ and type of coefficient. A, V, H and D stand for approximation,
vertical, horizontal and diagonal coefficients, respectively. The
scale given in column 4 is calculated 
as $1.5'(2^{J-j})$, where $2^J\times 2^J$ is the total number of
pixels in the map.
The total number of coefficients and those used in the analysis for each
region are given in the fifth and sixth columns. The rest of the
coefficients are discarded to avoid boundary effects.}
\label{tab:regions_mallat}
\begin{tabular}{|c|c|c|c|c|c|c|}
\hline
region & $j$ & detail & $\sim$scale($'$) & no.coeff. & used \\
\hline
1  & 1 & A & $\ge$384  & 4     & 0 \\
2  & 1 & V & 384 & 4     & 0  \\   
3  & 1 & H & 384 & 4     & 0  \\   
4  & 1 & D & 384 & 4     & 0  \\   
5  & 2 & V & 192 & 16    & 0   \\   
6  & 2 & H & 192 & 16    & 0   \\   
7  & 2 & D & 192 & 16    & 0   \\   
8  & 3 & V & 96 & 64    & 16  \\   
9  & 3 & H & 96 & 64    & 16  \\
10  & 3 & D & 96 & 64    & 16  \\   
11 & 4 & V & 48 & 256   & 144 \\   
12 & 4 & H & 48 & 256   & 144 \\   
13 & 4 & D & 48 & 256   & 144 \\   
14 & 5 & V & 24 & 1024  & 784 \\
15 & 5 & H & 24 & 1024  & 784 \\
16 & 5 & D & 24 & 1024  & 784 \\
17 & 6 & V & 12 & 4096  & 3481 \\
18 & 6 & H & 12 & 4096  & 3481 \\  
19 & 6 & D & 12 & 4096  & 3481 \\  
20 & 7 & V & 6 & 16384 & 14641 \\ 
21 & 7 & H & 6 & 16384 & 14641 \\ 
22 & 7 & D & 6 & 16384 & 14641 \\ 
23 & 8 & V & 3 & 65536 & 60025 \\ 
24 & 8 & H & 3 & 65536 & 60025 \\ 
25 & 8 & D & 3 & 65536 & 60025 \\ 
\hline
\end{tabular}
\end{table}

Fig.~\ref{fig:k4_gaus_cmb_z3} gives the wavelet statistic for the
input CMB map (top), the MEM (middle) and the WF (bottom)
reconstructions. All maps have been smoothed with a Gaussian beam with
fwhm=5$'$~arcmin. As expected, the value of the $\kappa_4$ statistics
for the input (Gaussian) map lie within the error bars, i.e., the map
is compatible with Gaussianity at all scales. We want to study if this
behaviour is preserved in the reconstructed maps. We find that both
MEM and WF produce very similar results and that the reconstructions
are still compatible with Gaussianity. If we compare the results for
the input and reconstructed maps, we find some small differences in
the values of the wavelet statistics at large $k$ (small scales),
which are mainly due to an imperfect reconstruction because of the
presence of instrumental noise. However, this effect is small and does
not modify the conclusion that the underlying CMB distribution is
consistent with Gaussian, according to this test.

Fig.~\ref{fig:k4_gaus_cmb_z3b} shows the results obtained from the
input (top) and reconstructed (bottom) combined maps, constructed by
combining the 143 and 217 GHz channels. The input combined map is
obtained as follows: first we create two maps containing only CMB,
filtered with the corresponding beams of the 143 and 217 GHz channels;
we then combine the two maps weighting them accordingly to the noise
levels of the two channels. This is the underlying CMB signal present
in the reconstructed combined map and therefore we should use it for
the comparison. Note that with this method, we can not aim to have
better results than those obtained for the input combined map, since
the reconstructed CMB signal is convolved and combined in that way.
Finally, both maps have been smoothed with a Gaussian beam of
fwhm=5~arcmin in order to increase the signal to noise ratio of the
reconstructed combined map. We find also in this case that the input
and reconstructed maps are compatible with Gaussianity. There are
again some small differences between both plots at high $k$'s which
are mainly due to the presence of instrumental noise in the
reconstruction.

Therefore, if the underlying CMB map is Gaussian, all the considered
methods preserve the Gaussian character of the temperature distribution.

\subsection{Non-Gaussian case}
\label{sec:ng}
In order to test the effect of component separation on the underlying
CMB temperature distribution, we have also considered two non-Gaussian
test maps. In particular, we have considered mixtures of Gaussian CMB
and cosmic strings in proportions 1:1 and 2:1 in rms. The proportion
2:1 corresponds approximately to the string contribution (18 per cent
in the power spectrum) of the best fit to the CMB power spectrum using
a mixture of inflation and cosmic strings found by \cite{bou01}.

%
%

\begin{figure}
\includegraphics[angle=0,width=\hsize]{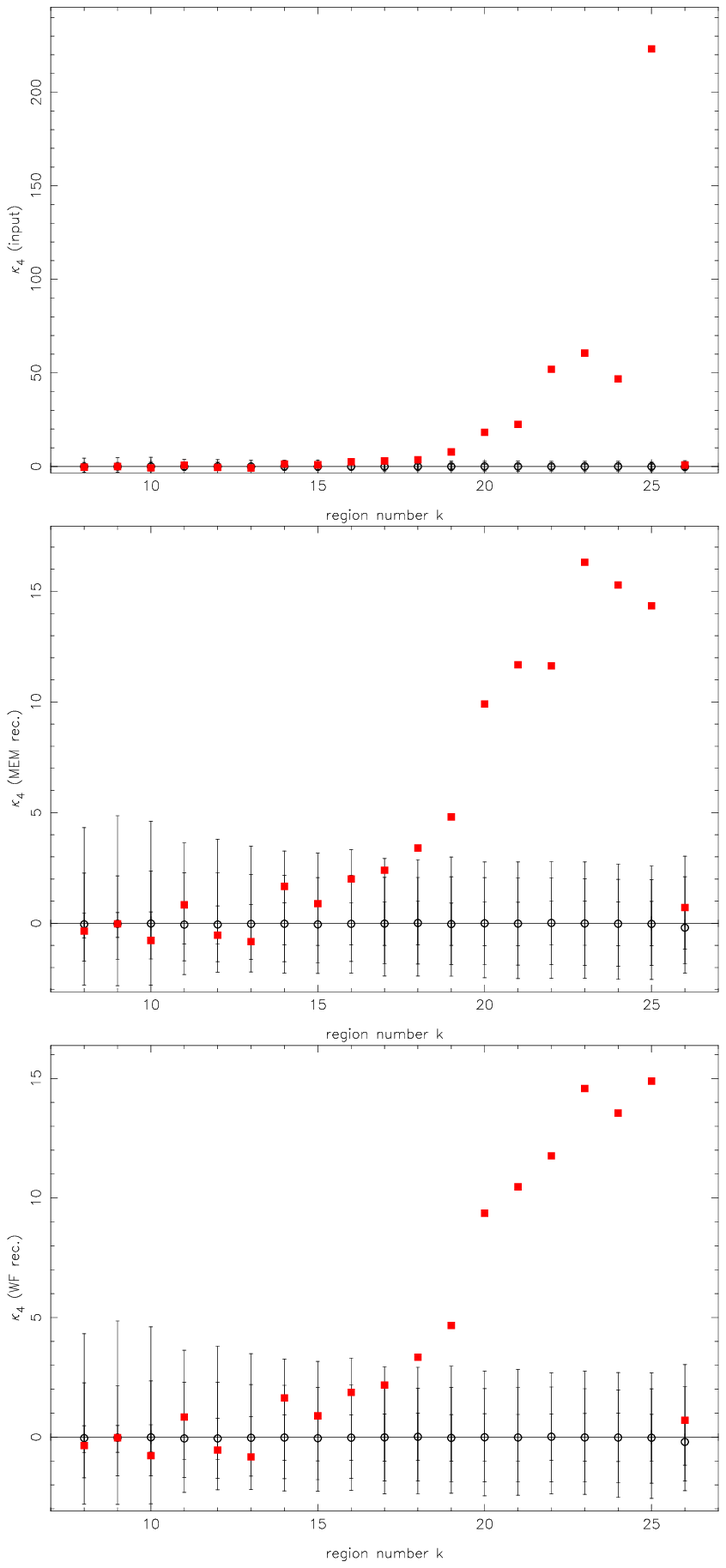}
\caption{The value of $\kappa_4$ versus the region number is shown for
the case of a mixture of Gaussian CMB and cosmic strings in proportion
1:1. The plots correspond to the input CMB (top), MEM reconstruction
(middle) and WF reconstruction (bottom).}
\label{fig:k4_mix11_cmb_z3}
\end{figure}

For the non-Gaussian case with proportion 1:1, we show in
Fig.~\ref{fig:k4_mix11_cmb_z3}, the wavelet statistics for
the 5 arcmin smoothed input (top), MEM (middle) and WF (bottom) maps.
For the input map, we find very clear non-Gaussian detections for large $k$,
greater than ~200$\sigma$ at $k=25$.
For the MEM reconstruction, the detections are also very clear (up to
$\sim 16.5 \sigma$) but they have been significantly lowered with respect to
the input map. Therefore there is a clear bias in the values of
$\kappa_4$, and the CMB reconstruction tends to be more Gaussian than
the input one. This is due to an imperfect reconstruction and loss of
resolution caused mainly by the presence of instrumental noise.
At those scales where the signal-to-noise ratio is low, the entropic
prior tends to produce a conservative reconstruction with little
structure, which causes the temperature distribution to be closer to
Gaussian.  A similar result is found for the WF reconstruction, where
the highest non-Gaussian detection is at the level of $\sim 15
\sigma$.  Since WF assumes a Gaussian prior for the signal to be
recovered, this tends to make the reconstruction even more Gaussian
than does the entropic prior.
\begin{figure}
\includegraphics[angle=0,width=\hsize]{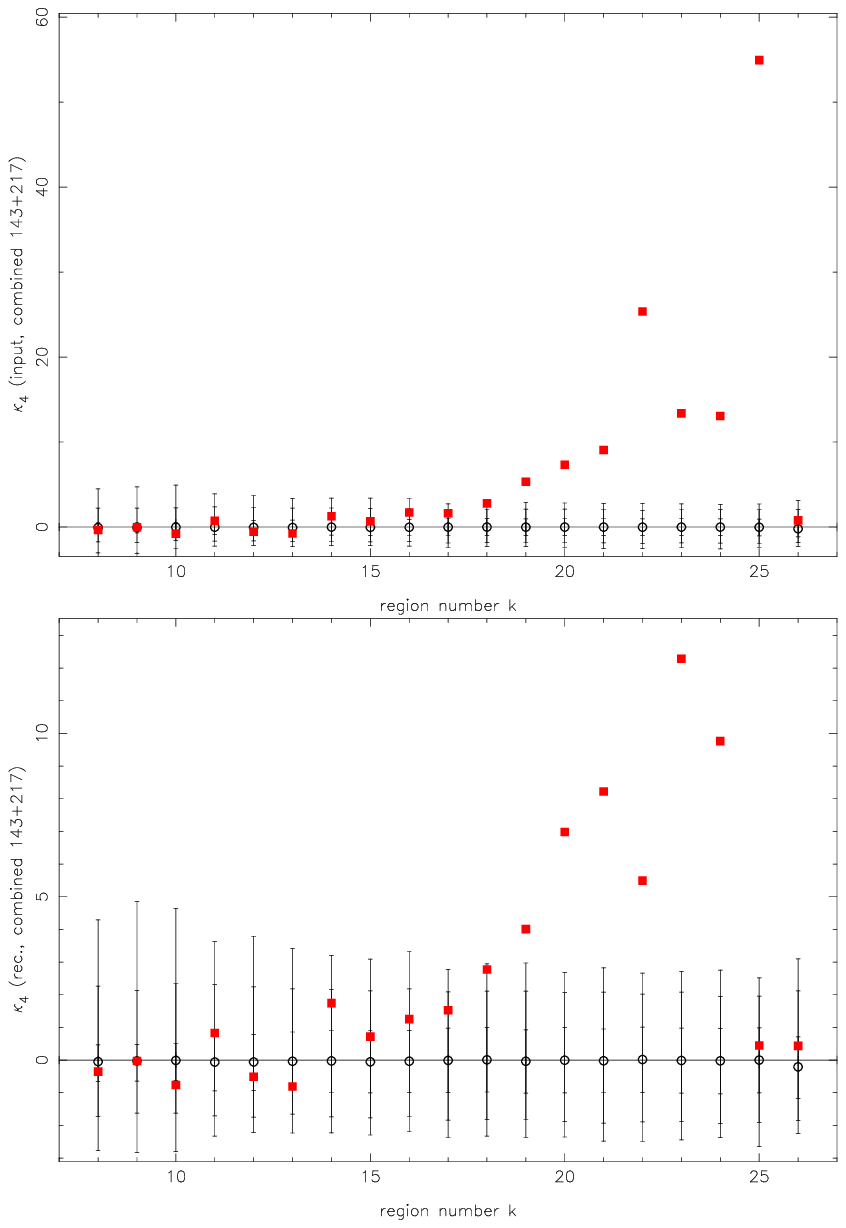}
\caption{The value of $\kappa_4$ for the input (top) and reconstructed
(bottom) combined maps (143+217 GHz) for the case of a mixture of
Gaussian CMB and cosmic strings in proportion 1:1.}
\label{fig:k4_mix11_cmb_z3b}
\end{figure}
Fig.~\ref{fig:k4_mix11_cmb_z3b} shows the results for the combined
method. The non-Gaussianity is detected at the input combined map at a
very high level ($\la 55 \sigma$). In the reconstructed combined map,
the non-Gaussianity is also clearly detected but at a lower level
($\la 12\sigma$). Therefore, the reconstructed map is also more
Gaussian than the input one, mainly due to the contamination coming
from instrumental noise. WF and MEM produce quite similar results,
although, in general, the MEM reconstruction produces slightly higher
detections than WF.  Also, the level of the detections obtained with
the LCFC are lower than those of the MEM and WF reconstructions. Using
different noise realisations only produces a small dispersion in the
detections and does not modify the results.

\begin{figure*}
\includegraphics[angle=0,width=\hsize]{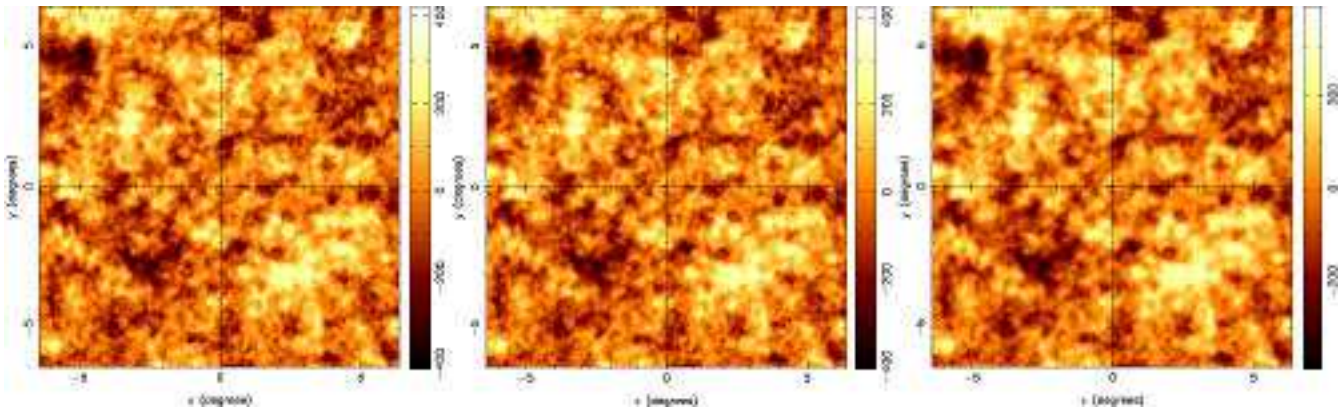}
\caption{Reconstructed CMB map obtained with MEM (left), WF (middle)
and a linear combination of the 143 and 217 GHz channels (right) from
Planck simulated data using as input a mixture of Gaussian CMB and
cosmic strings in proportion 2:1. The
maps have been smoothed with a Gaussian beam of fwhm=5 arcmin. The
units are thermodynamic temperature in $\mu$K.} 
\label{fig:rec_mix21_cmb_z3}
\end{figure*}

\begin{figure}
\includegraphics[angle=270,width=\hsize]{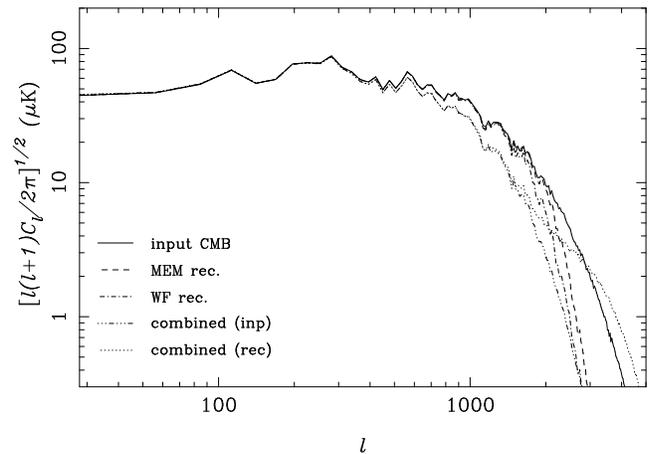}
\caption{Power spectra obtained from the input and reconstructed CMB
  maps for the non-Gaussian case with a mixture of the Gaussian CMB
  and cosmic strings in proportion 2:1. The maps have been smoothed
  with a Gaussian beam of fwhm=5 arcmin.}
\label{fig:ps_mix21_cmb_z3}
\end{figure}

We have also considered map with a lower non-Gaussianity level
constructed by mixing the Gaussian CMB map with the cosmic strings map
in the proportion 2:1 in rms (right panel of
Fig.~\ref{fig:input_cmb}). The reconstructions obtained using MEM, WF
and the LCFC are given in Fig.~\ref{fig:rec_mix21_cmb_z3} (smoothed
with a Gaussian beam of fwhm=5~arcmin). Note that again the MEM and WF
reconstructions are quite similar, whereas the LCFC has lower
resolution.
The corresponding reconstructed and input power spectra are also given
in Fig.~\ref{fig:ps_mix21_cmb_z3}. As before, MEM is able to recover
the high $\ell$'s better than WF, whereas the LCFC presents an excess
of power at high multipoles due to the presence of instrumental noise.

\begin{figure}
\includegraphics[angle=0,width=\hsize]{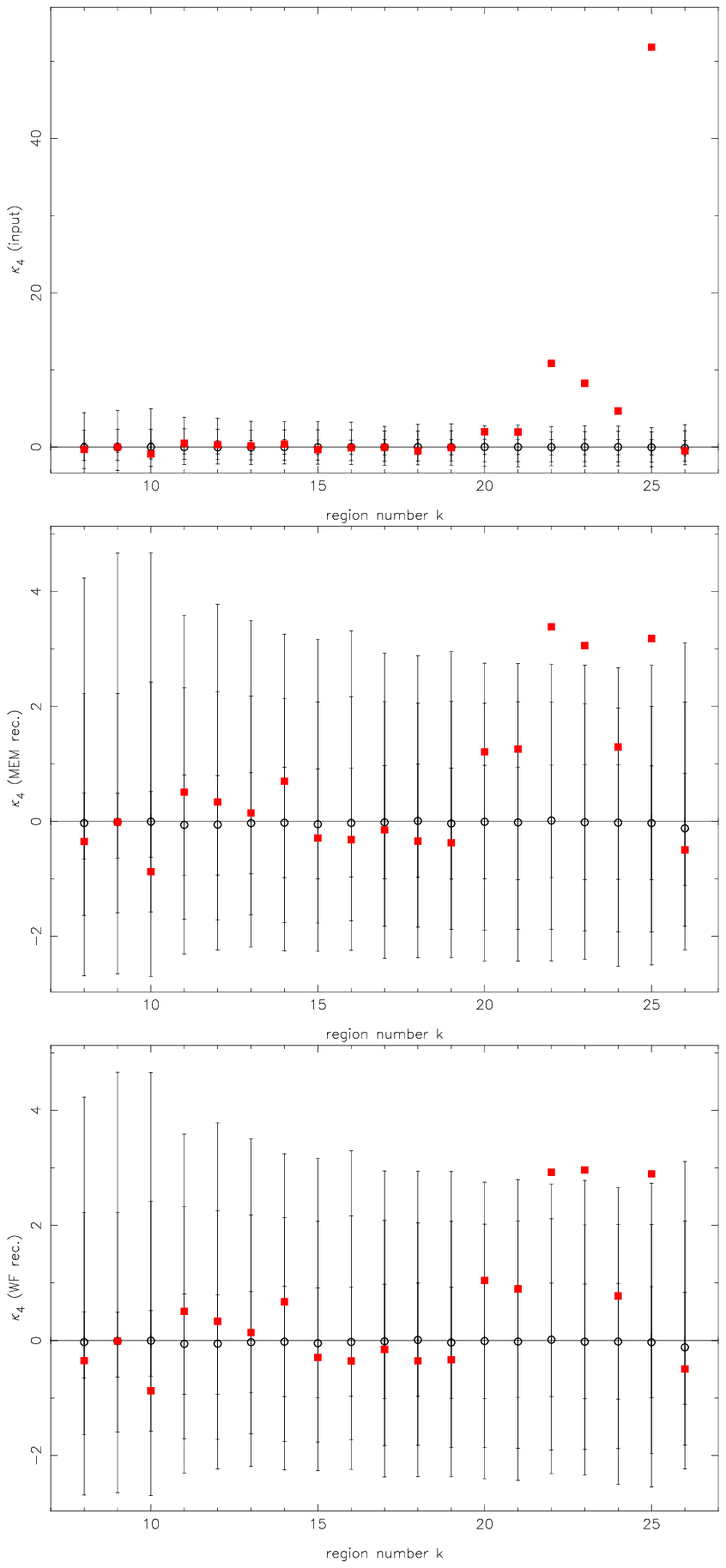}
\caption{Values of $\kappa_4$ at each wavelet region for
the case of a mixture of Gaussian CMB and cosmic strings in proportion
2:1: input CMB (top), MEM reconstruction
(middle) and WF reconstruction (bottom).}
\label{fig:k4_mix21_cmb_z3}
\end{figure}

Fig.~\ref{fig:k4_mix21_cmb_z3} shows the value of $\kappa_4$ versus
the region number for the input (top), MEM (middle) and WF (bottom)
maps for this second non-Gaussian test map. The non-Gaussianity is
still clearly detected in the input map although, as expected, the
number and level of detections are lower than for the mixture in
proportion 1:1. In particular there are 4 detections of
non-Gaussianity outside the 99 per cent acceptance region, with the
highest detection found at $k=25$ at the level of $\sim 50 \sigma$. In
the reconstructed maps, the level of the detections decrease
dramatically. For MEM, we find 3 non-Gaussian detections at level $\la
3.5 \sigma$, whereas for WF there are the same 3 detections but at
slightly lower level ($ \la 3\sigma$).  Therefore there is again a
large bias in the sense of gaussianising the reconstruction. Since we
are in the limit of the detection, different noise realisations can
make that the detections fall inside the 99 per cent region of
acceptance. In any case, we find again that the MEM reconstruction
produces in general slightly higher detections than the one of WF.

\begin{figure}
\includegraphics[angle=0,width=\hsize]{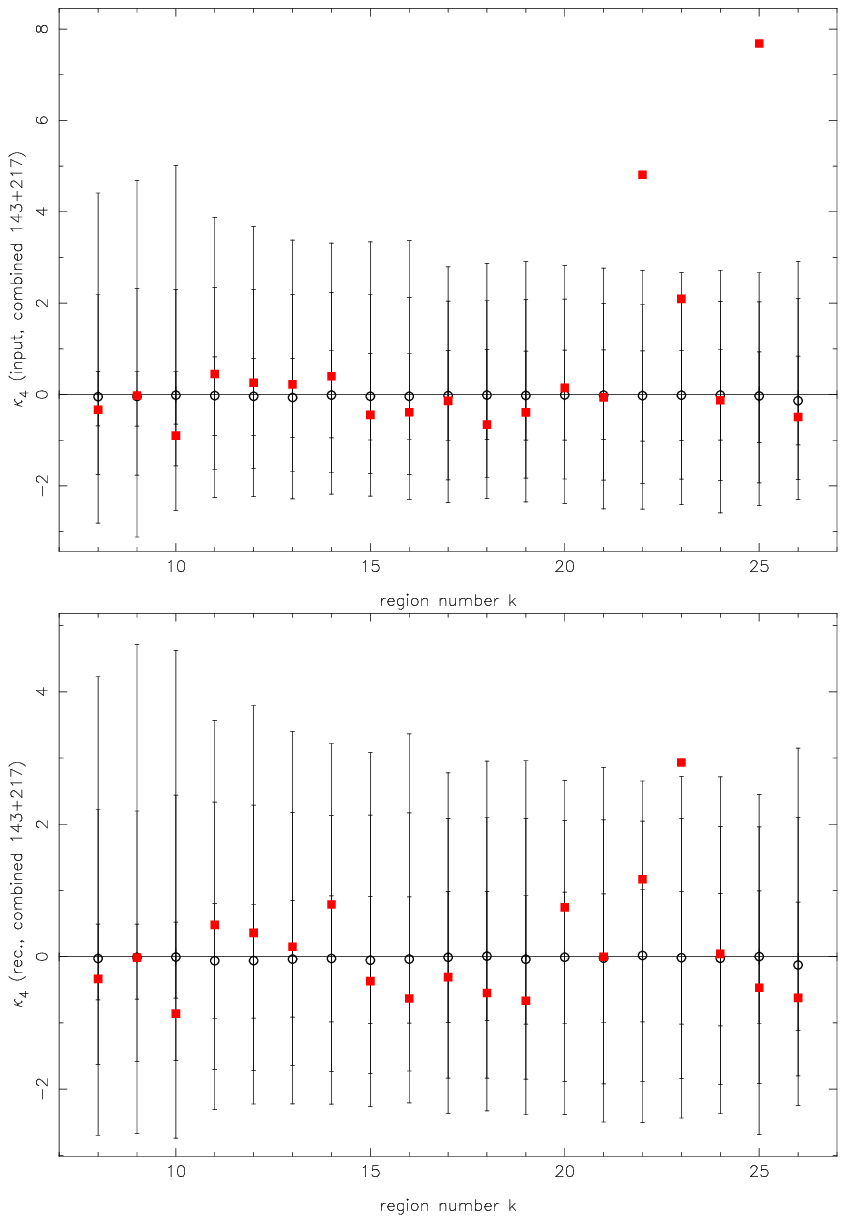}
\caption{The values of $\kappa_4$ versus the wavelet region are given
for the input (top) and reconstructed (bottom) combined (143+217 GHz)
maps corresponding to the case of a non-Gaussian initial CMB generated
as the mixture of Gaussian CMB and cosmic strings in proportion 2:1.}
\label{fig:k4_mix21_cmb_z3b}
\end{figure}

With regard to the LCFC, the results are given in
Fig.~\ref{fig:k4_mix21_cmb_z3b}. Non-Gaussianity is detected at the
$\sim 8 \sigma$ level in the input combined map. The reconstructed map
shows a detection outside the 99 per cent acceptance region at region
$k=23$. However, this detection was present in the input only outside
the 95 per cent acceptance region. This result is a combination of
having a significant intrinsic $\kappa_4$ value and the noise
realisation and thus is not a robust detection. In fact, when using a
different noise realisation, this detection disappears and other
detections can appear. Therefore, we are again in the limit of
detecting the non-Gaussianity. In any case, the levels of $\kappa_4$
are lower than those in the MEM and WF reconstructions.

Thus, in all the considered cases, there is a clear bias for the three
studied methods and the intrinsic non-Gaussianity is strongly reduced in the
reconstructed map. 

\subsection{Effect of point sources}
\label{sec:point_sources}
For simplicity, we have not included the emission of extragalactic
point sources in the previous subsections. Now, we will include this
contaminant in the simulations and study the reconstructions
obtained using only MEM and the MEM+MHW joint analysis.
We use the Gaussian CMB map given in the left panel of
Fig.~\ref{fig:input_cmb} as input.
\begin{figure}
\begin{center}
\includegraphics[angle=0,width=5.8cm]{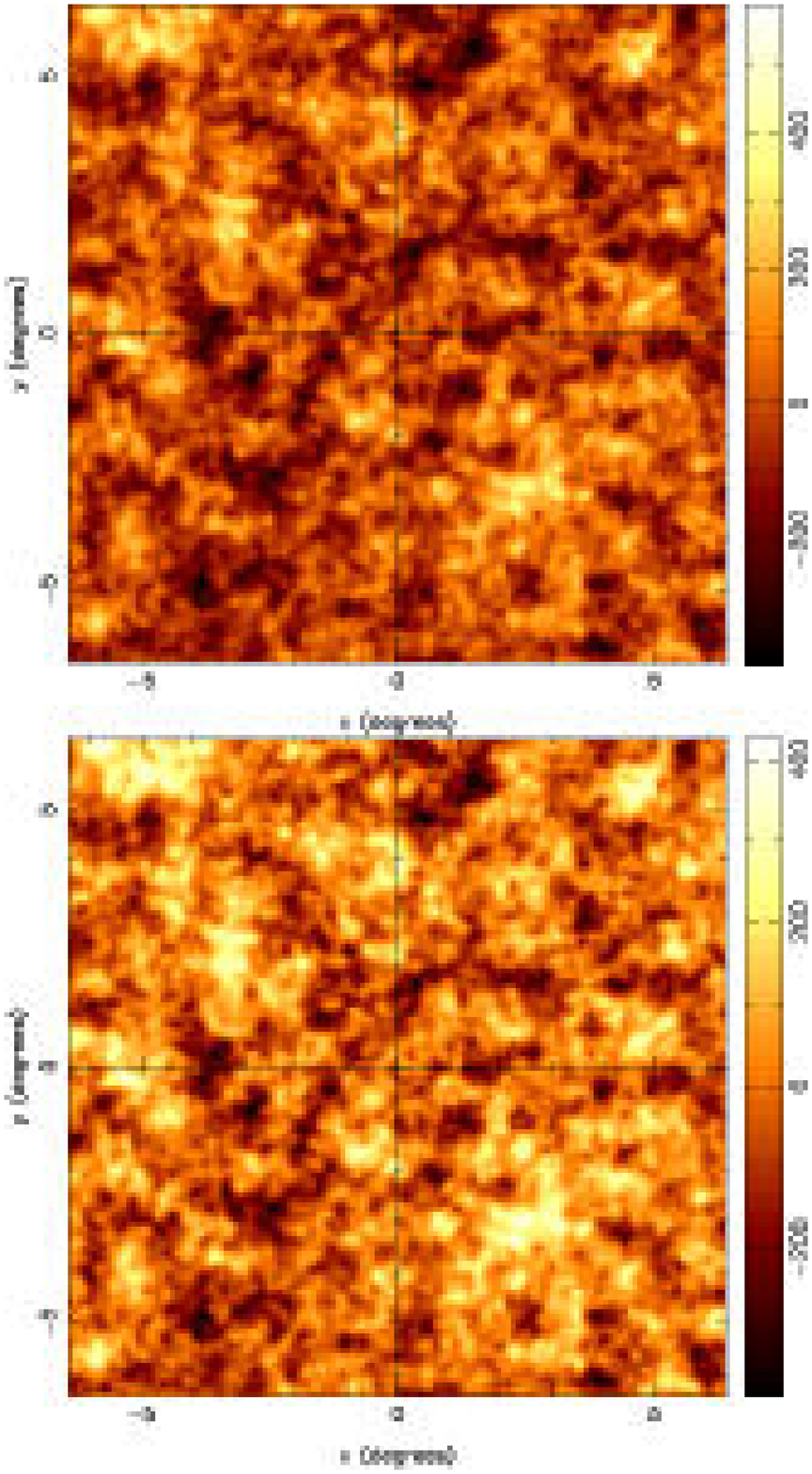}
\caption{Reconstructed CMB map obtained with MEM (top) and MEM+MHW
(bottom) for the case when point sources are included in the data. The
reconstructions should be compared with the input Gaussian CMB (left
panel of Fig.~\ref{fig:input_cmb}.
The maps have been smoothed with a Gaussian beam of fwhm=5 arcmin and the
units are thermodynamic temperature in $\mu$K.} 
\label{fig:rec_sources_cmb_z3}
\end{center}
\end{figure}
\begin{figure}
\includegraphics[angle=0,width=\hsize]{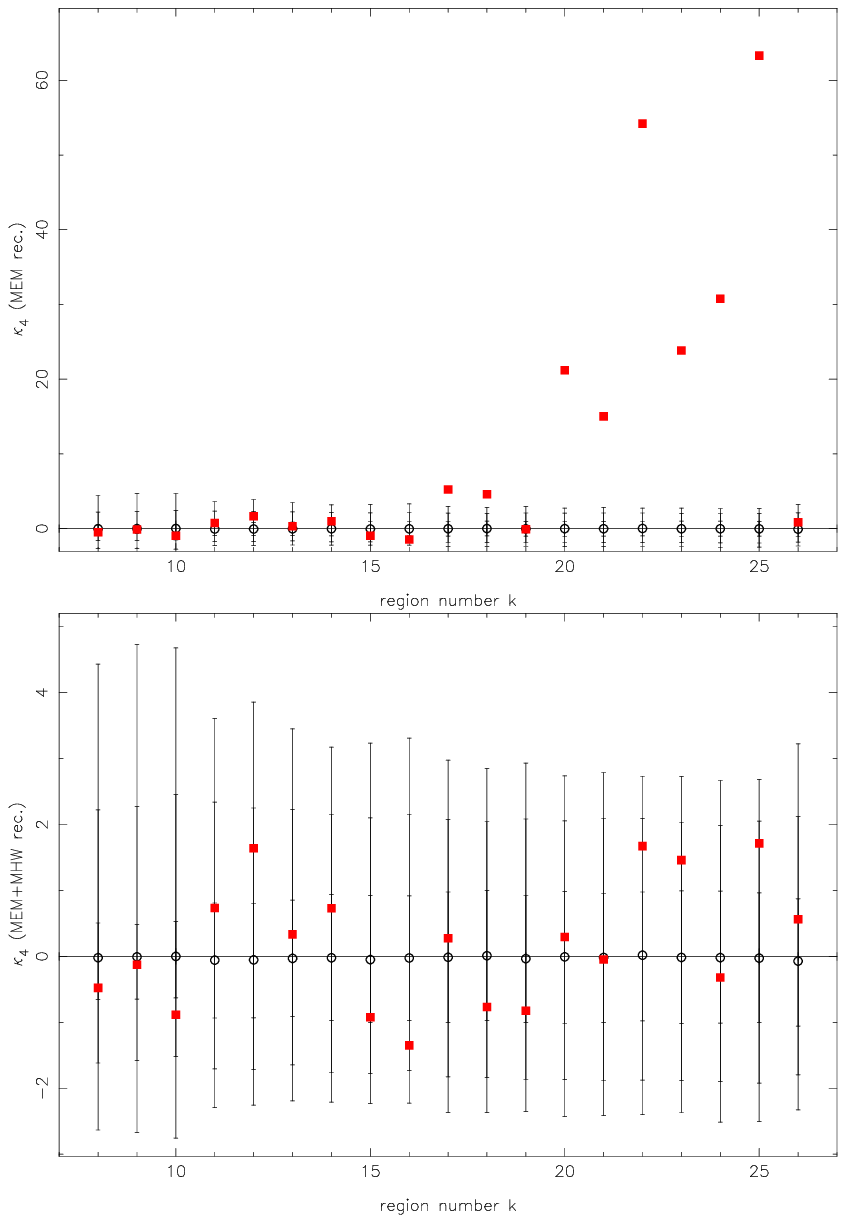}
\caption{Values of $\kappa_4$ for each wavelet region for the CMB
reconstruction obtained using only MEM (top) and the MEM+MHW joint
analysis (bottom) when point source emission is included in the
data for an initial Gaussian CMB. The results should be compared with
the top panel of Fig.~\ref{fig:k4_gaus_cmb_z3}, which shows the values of
$\kappa_4$ for the input CMB map.}
\label{fig:k4_fuentes_z3}
\end{figure}

Fig.~\ref{fig:rec_sources_cmb_z3} gives the reconstructions obtained
using only MEM (top) and the MEM+MHW joint analysis (bottom) smoothed
with a Gaussian beam of fwhm=5 arcmin. The recovered power spectra for
both cases is very similar to the one obtained for the MEM
reconstruction of Fig.~\ref{fig:ps_gaus_cmb_z3} (when point source
emission was not included). However, even if not seen in the power
spectrum, the MEM reconstruction is clearly contaminated by point
source emission (see top panel of
Fig.~\ref{fig:rec_sources_cmb_z3}). This contamination is reduced when
using the joint method (bottom panel of
Fig.~\ref{fig:rec_sources_cmb_z3}). This effect is also seen in
Fig.~\ref{fig:k4_fuentes_z3}, which shows the corresponding wavelet
statistic for the previous reconstructed maps. These results should be
compared with the top panel of Fig.~\ref{fig:k4_gaus_cmb_z3}, which
shows the values of $\kappa_4$ for the input CMB map. We find clear
detections of non-Gaussianity in the MEM reconstructed map, at the
level $\la 60 \sigma$. Since the underlying CMB signal is Gaussian,
this is due to the presence of residual point source emission, which
MEM has not been able to remove. However, when the MHW is used to
subtract the brightest point sources prior to the application of MEM,
this contamination is greatly reduced and the CMB reconstruction is
again compatible with Gaussianity (bottom panel of
Fig.~\ref{fig:k4_fuentes_z3}).  Similar results have been found when
combining the MHW with WF or with the LCFC.  If not subtraction of
point sources is attempted, both methods detect residual point
sources. In particular, for WF this detection
is slightly lower than for MEM (since it tends to gaussianise more
the residuals) whereas for the LCFC it is also reduced (due to the lower
resolution of the combined map). But if the MHW is previously
applied, the CMB reconstruction is also compatible with Gaussianity.
Therefore, the removal of point source emission is crucial in order to
avoid the introduction of spurious non-Gaussianity in the
reconstructed CMB temperature distribution.

As a further test, we have also applied the MEM+MHW joint analysis to
Planck simulated data containing point sources and a non-Gaussian
input CMB. In this case, we find that the level of the detection of   
non-Gaussianity is similar to that found in section~\ref{sec:ng}
for the two considered proportions of cosmic strings.

\section{Role of instrumental noise and foregrounds}
\label{sec:role}


In order to test which is the effect of each of the elements that take
part in the component separation method, we present in this section
some additional tests. In particular, we perform first some analyses to
understand the effect of the Galactic foregrounds and the SZ
emissions, without including the contribution of point sources,
on the CMB reconstructed distribution. Then, we consider the same case
but reducing the level of noise by a factor of 10 and, finally, we
further study the effect of including point sources.

\subsection{Galactic foregrounds and SZ emission}
\label{sec:foregrounds}

\begin{figure}
\includegraphics[angle=0,width=\hsize]{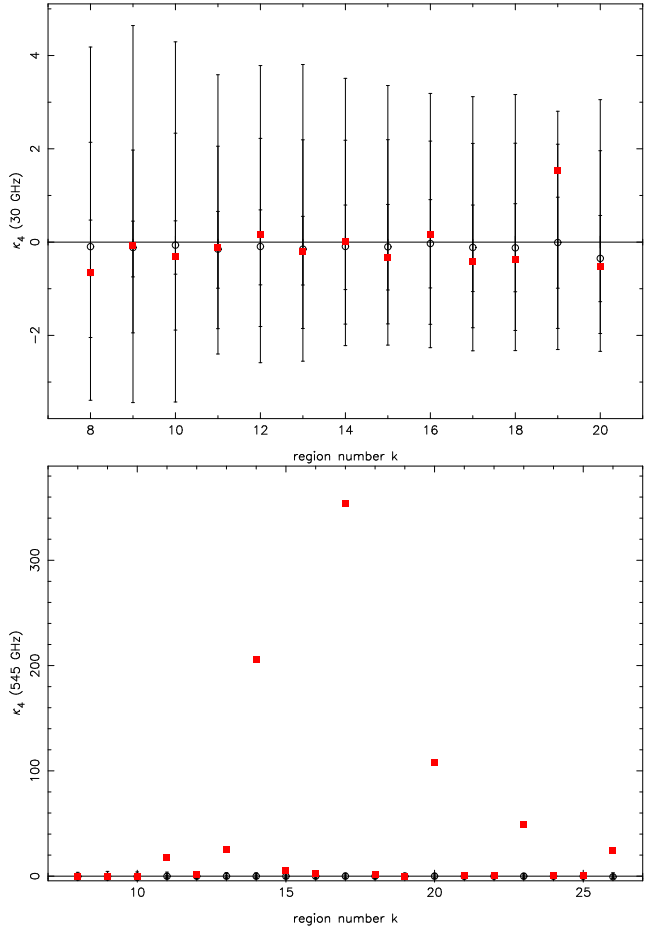}
\caption{Values of $\kappa_4$ for each wavelet region for the 30 GHz
(top) and the 545 GHz (bottom) frequency channels simulated using the
Galactic foregrounds given in Fig.~\ref{fig:foregrounds}. The 30 and
545 frequency channels have been smoothed with a Gaussian beam of 20
and 5 arcminutes respectively to reduce the instrumental noise. Note
that since the 30 GHz channel has a pixel of 6 arcminutes, only
wavelet regions with $k \le 19$ are considered, whereas $k=20$
corresponds in this plot to the value obtained from the real map.}
\label{fig:k4_fore_z3}
\end{figure}

First of all, we have applied our Gaussianity analysis to a low and a
high frequency Planck channel (where the synchrotron and dust
emissions are more important) without including point source emission
in order to show how the wavelet coefficients are affected by the
presence of foregrounds. Fig.~\ref{fig:k4_fore_z3} gives the values of
$\kappa_4$ for the 30 GHz (top) and 545 GHz (bottom) channels
including the Galactic foregrounds of Fig.~\ref{fig:foregrounds} as
well as our reference Gaussian CMB and the thermal and kinetic SZ
emissions (the results are qualitatively similar for the other three
considered Galactic regions). Since the pixel size at 30 GHz is 6
arcmin, we consider wavelet regions only up to $k=19$, whereas $k=20$
corresponds to the $\kappa_4$ value of the real map. We see that, at
this frequency, there are not deviations from Gaussianity in spite of
having a certain contamination from synchrotron. At high Galactic
latitude, this emission is subdominant in comparison with the CMB and
it is not seen in this test. We also have to point out that the most
important contribution of the synchrotron is expected to occur at
larger scales than those considered here, therefore the synchrotron
could be important when analysing larger regions of the sky. Finally,
one should bear in mind that the synchrotron template used in this work
is based on the Haslam map, which has a resolution of $0.85^\circ$ and
therefore it does not contain real structure (only simulated) at the
smallest scales, which may not be representative of the true
emission. However, even with these uncertainties, we do not expect the
results of this work to change with a more realistic simulation of
synchrotron. This is so because the reconstructions obtained from MEM
and WF were very robust under variations of the Galactic foregrounds
and, with regard to the LCFC, the synchrotron is negligible at the
considered frequency channels (143 and 217 GHz). Regarding the 545 GHz
channel, which is completely dominated by dust emission, we see very
clear non-Gaussian detections at different scales (bottom panel of
Fig.~\ref{fig:k4_fore_z3}). Non-Gaussian detections at different
levels are also found at 353 GHz for the four considered Galactic
regions whereas at 217 GHz there is one detection of non-Gaussianity
only for the brightest region (region 1). Therefore, we see that the
dust template is strongly non-Gaussian but for frequencies of 217 GHz
or lower the dust emission seems to be, in general, already quite weak
and should not affect our Gaussianity analysis (although, of course,
it could be more important at larger scales than those considered
here). Of course, this is assuming that the uncertainties on the simulated
dust emission are small.

To further test the effect of foregrounds we have repeated the
analyses of sections~\ref{sec:gaussian} and~\ref{sec:ng} using data
maps where the foreground components (Galactic emission and SZ) have
been divided by a factor of 100. In this case the LCFC is done
directly using these frequency maps (without further reducing the
Galactic foregrounds). The results are very similar to those obtained
in the previous section for both the Gaussian and non-Gaussian
cases. These results confirm that, for the wavelet analysis presented
in this work and within the uncertainties of our simulations, the
Galactic foregrounds (for high Galactic latitude regions) and SZ
emissions do not seem to introduce any spurious non-Gaussianity.

\subsection{Instrumental noise}
\label{sec:noise}

\begin{figure}
\includegraphics[angle=0,width=\hsize]{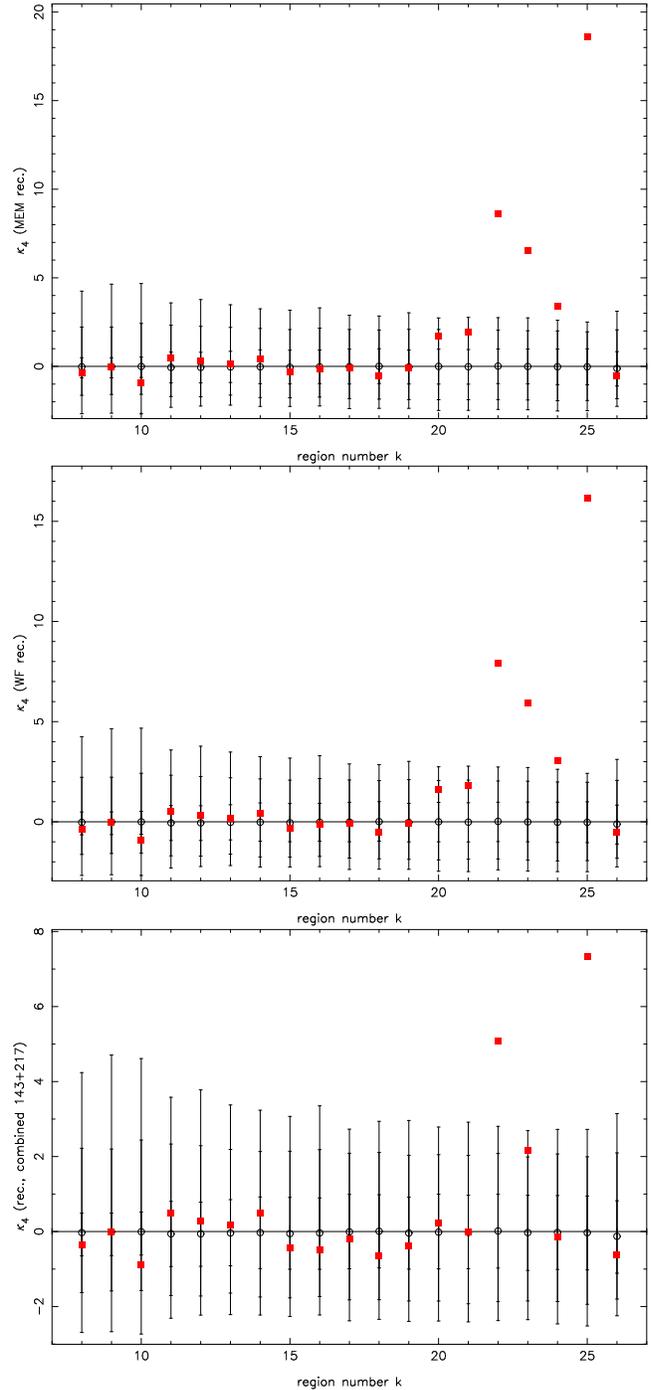}
\caption{Values of $\kappa_4$ at each wavelet region for the case of a
mixture of Gaussian CMB and cosmic strings in proportion 2:1 where the
noise has been reduced by a factor of 10. The different panels correspond
to MEM reconstruction (top), WF reconstruction (middle) and 143+217
GHz combined map (bottom) smoothed with a Gaussian beam of fwhm=5 arcminutes.}
\label{fig:k4_mix21_cmb_z3_low10}
\end{figure}

To understand the effect of instrumental noise on the CMB
reconstructions, we have repeated the Gaussian analyses of
sections~\ref{sec:gaussian} and~\ref{sec:ng} reducing the noise dispersion
by a factor of 10 (and again without including point sources). When
the input CMB is Gaussian, the reconstructed CMB is again Gaussian for
MEM, WF and LCFC. The main difference is that the values of $\kappa_4$ of
the reconstructed maps are closer to the same values in the input
map. This is especially visible at the smallest scales (largest values
of $k$) which were more affected by the presence of noise.  With
regard to the non-Gaussian case, the detections are clearly improved
for the three methods and for the two proportions of cosmic strings
considered. Fig.~\ref{fig:k4_mix21_cmb_z3_low10} shows the $\kappa_4$
statistics for the reconstructed CMB map using MEM (top), WF (middle)
and the combined 143+217 GHz map (bottom) for a mixture of Gaussian
CMB and strings in proportion 2:1. This figure should be compared with
Fig.~\ref{fig:k4_mix21_cmb_z3} and~\ref{fig:k4_mix21_cmb_z3b}. It
becomes apparent that lowering the noise allows a much better
detection of the intrinsic non-Gaussianity in all the
cases. Therefore, the instrumental noise is clearly impairing our
ability to detect intrinsic non-Gaussianity. This also indicates that
MEM and WF perform better than the simple LCFC not only because they
remove better the foregrounds but mainly because they attempt to
denoise (and deconvolve) the signal.

\subsection{Point sources}
\label{sec:point_sources_low10}

\begin{figure}
\includegraphics[angle=0,width=\hsize]{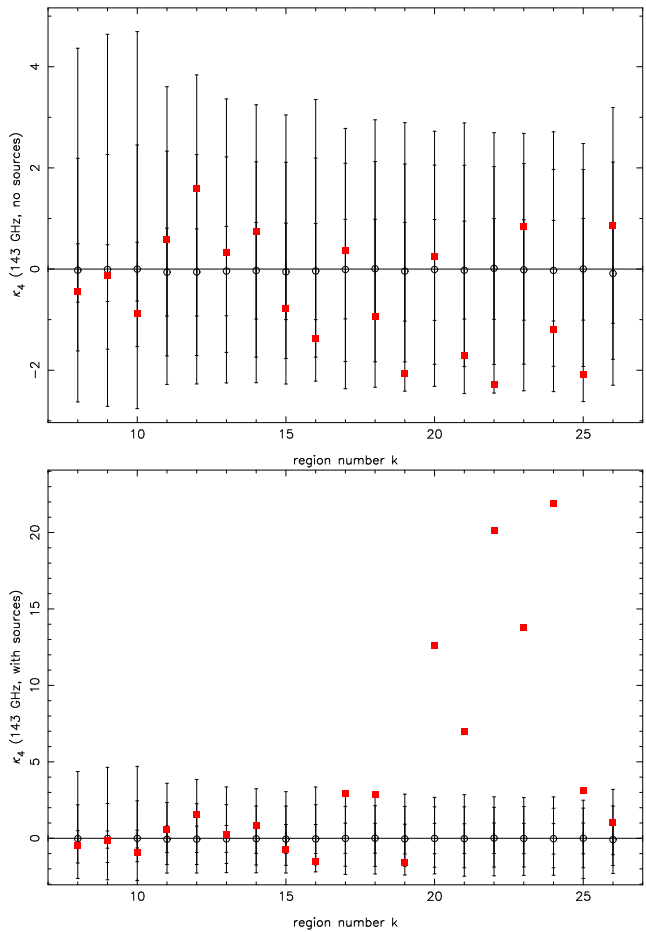}
\caption{Values of $\kappa_4$ at each wavelet region for the 143 GHz
channel without including (top panel) and including the emission of
point sources (bottom panel). In order to reduce the noise, both maps
have been smoothed with a Gaussian beam with fwhm=5 arcminutes.}
\label{fig:k4_c143}
\end{figure}

In section~\ref{sec:point_sources} we have showed that point sources
can be a very damaging foreground and that is crucial to remove, at
least, the brightest point sources in order to avoid the presence of
spurious non-Gaussianity in the CMB distribution. As an example of the
effect of extragalactic point sources, we show in
Fig.~\ref{fig:k4_c143} the $\kappa_4$ statistics for the 143 GHz
channel without including (top) and including (bottom) point source
emission. It is clear that point sources are introducing important
levels of non-Gaussianity in the data, especially at the smallest
scales, even for the 143 GHz channel where the contribution of point sources
is expected to be relatively small.
%

\begin{figure}
\includegraphics[angle=0,width=\hsize]{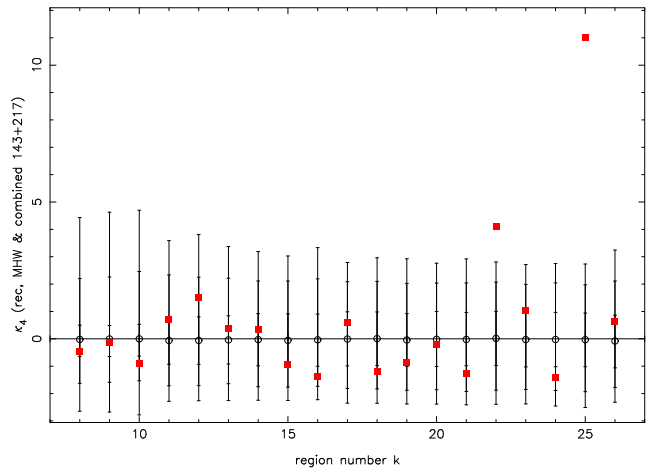}
\caption{Values of $\kappa_4$ at each wavelet region for the 143 and
217 GHz combined map including point source emission for a case where
the instrumental noise has been reduced by a factor of 10. The
combined map has been smoothed with a Gaussian beam with fwhm=5
arcminutes. Before combining the maps, the brightest point sources
have been removed using the Mexican Hat wavelet.}
\label{fig:k4_sources_low10}
\end{figure}

We would like to point out that we are performing our non-Gaussian
analysis on a small patch of the sky of $12.8^\circ \times
12.8^\circ$. However Planck will provide all-sky observations of the
CMB and therefore we expect to be able to analyse much larger regions
of the sky. Since having a larger area of the sky will appreciably
reduce the variance of the $\kappa_4$ statistics we may wonder if, in
this case, we would be able to detect the point source residuals even
after subtracting the brightest point sources. We have tried to give
an answer to this question with the following qualitative
argument. For the LCFC, the part of the variance coming from Gaussian
white noise will decrease as the considered area. For two-thirds of
the sky this corresponds to a reduction in the part of the dispersion
coming from the noise of the order of 10. Therefore, in order to mimic
the case of performing the component separation in a larger area of the sky,
we have constructed new data reducing the noise dispersion by a factor
of 10 and subtracted from each channel the same catalogue of point
sources as that of section~\ref{sec:point_sources}. Then we have
applied our non-Gaussianity test on the map obtained with the LCFC
using the 143 and 217 GHz channel. The result is given in
Fig.~\ref{fig:k4_sources_low10}, showing a clear detection of
non-Gaussianity. Of course, in the real case, the error bars will be
further reduced, although not in such a straightforward way as for the
white noise, since we are not considering the reduction of the variance
corresponding to the CMB and the point sources themselves. But this means that
our estimation can be considered as an upper limit to the error bars
of the $\kappa_4$ and thus the residual point source emission will be
even more clearly detected.

For the case of MEM, given that is a non-linear process, the way in
which the dispersion of the $\kappa_4$ statistics scales with the
area is not so simple, but we still expect our argument to hold
qualitatively. We have performed the Gaussian analysis for the data
with the noise level reduced by a factor of 10 using MEM+MHW and found
again clear detections of non-Gaussianity due to the residual point
sources. 

Therefore, point source emission should be carefully reduced or masked
on CMB data since otherwise it will introduce spurious non-Gaussianity on the
analysed signal.

\section{Discussion}
\label{sec:discussion}
We have studied how different component separation techniques affect
the underlying CMB temperature distribution. In particular, we have
used Planck simulated data (including the cosmological signal,
Galactic foregrounds, SZ effects and instrumental noise) to
reconstruct the CMB with MEM, WF and LCFC.  We have also studied the
effect that point source emission has in the reconstructed map
obtained from MEM and a MEM+MHW joint analysis. The impact of the
component separation method on the CMB has been quantified by
performing a wavelet analysis of the reconstructed CMB map and
compared it with the same analysis for the input map. Gaussian and
non-Gaussian CMB simulations have been tested. We have also presented
some additional tests to understand the effect of the different
foregrounds and instrumental noise.

Using as input a Gaussian CMB map (and without including the emission
from point sources in the data), the three considered methods lead to
a CMB reconstructed map which is compatible with
Gaussianity. Therefore, none of them introduces spurious
non-Gaussianity in the underlying CMB temperature distribution.
However, when the input CMB distribution is non-Gaussian, all the
methods introduce a strong bias in the reconstructed map, which
clearly reduces the level of the intrinsic non-Gaussian detections
when compared with the input map. As has been shown in
section~\ref{sec:noise}, this effect is mainly due to the presence of
instrumental noise which leads to an imperfect reconstruction.

WF assumes a Gaussian prior for the signal to be reconstructed and
therefore it is expected to perform well when the underlying CMB
distribution is Gaussian.  It is also expected that MEM will perform
at least as well as WF since it has been shown that the entropic prior
tends to the Gaussian one in the small fluctuation limit
\citep{hob98}. Thus, it is not surprising that both methods perform
equally well when the underlying CMB is Gaussian. However, for
non-Gaussian fields MEM would tend to perform better than WF. This is
the case for our simulations since we find that the level of detection
of intrinsic non-Gaussianity is, in general, slightly higher for the
MEM reconstructions than for the ones obtained with WF.  Although for
the considered cases the differences between both reconstructions are
small, they could be more important for other type of
non-Gaussianity. For instance, \cite{hob98} also show that the thermal
SZ cluster profiles reconstructed with MEM are closer to the true ones
than those obtained with WF.  It is also interesting to note that,
provided that we assume a good knowledge of the spectral behaviour of
the Galactic foregrounds, their effect in the MEM and WF
reconstructions is quite small. In fact, we have obtained our results
for four different sets of Galactic foregrounds and found that the CMB
reconstructions are quite insensitive to the input foreground maps.
Regarding the LCFC technique, it produces lower non-Gaussian
detections levels than both MEM and WF. We also find that, provided
that we choose frequency channels which are relatively clean from
Galactic foregrounds, the results do not depend on the chosen Galactic
region (outside a Galactic cut with $|b| = 20^\circ$). This result is
confirmed by the tests performed in section~\ref{sec:foregrounds},
where only the thermal dust seems to be playing a role at the highest
frequency channels. However, one should bear in mind that we are
considering only small patches of the sky and that the foregrounds are
expected to be more important at large scales. In addition, our
simulated templates are subject to uncertainties that can be
especially important at subdegree scales. We should also point out
that, although our wavelet technique has not been able to pick the
presence of foreground residuals, other type of studies (e.g. phase
analyses) may be more sensitive to the presence of foreground
contamination. 

The fact that MEM and WF provide better reconstructions than the LCFC
is expected since MEM and WF are very powerful methods which make use
of all the information present in the data to separate all the
components. In addition, they try to deconvolve and to denoise the
signal, what clearly improves our ability to recover the CMB. However,
they require some important assumptions which can make the method less
robust. On the contrary, the assumptions required by the LCFC are less
strong. When using MEM and WF, the frequency dependence of all the
components is assumed to be known. This is valid for the CMB, kinetic
and thermal SZ effects but is only approximated for the Galactic
foregrounds. Moreover the Galactic foregrounds are assumed to have a
pixel independent spectral index, which may be a reasonable
approximation for small patches of the sky (as the ones considered
here) but it does not hold when considering large fractions of the
sky. In addition, the results previously presented have been obtained
assuming full knowledge of the power spectra and cross-correlations
between the different components for both MEM and WF, although this
assumption can be relaxed.

For small patches of the sky, one could try to obtain directly from
the data the spectral behaviour of the foregrounds (assuming that it
is spatially invariant). This could be done, for instance, by
applying a blind-source separation method. Then this information could
be included in the MEM (or WF) algorithm to perform the final
reconstruction. Another solution is to use the extension of MEM
recently developed by \cite{sto05}, which successfully accommodates
spatially dust temperature variations. They find that the quality of
the CMB reconstruction is comparable to the one obtained in the ideal
case, when the dust temperature was assumed to be constant over the
sky. The method can also be used to deal with other foregrounds.

With regard to the influence of the assumed power spectra, we have
repeated the analysis using MEM assuming neither knowledge of the
power spectra nor of the cross-correlations between the
components. When the input temperature distribution is Gaussian, the
reconstructed CMB map is again compatible with Gaussianity. For the
non-Gaussian cases, the level of the detections are decreased. For the
map with Gaussian:non-Gaussian proportion of 1:1, the highest
non-Gaussian detection goes down to $\sim 10 \sigma$ (versus $\sim
16.5 \sigma$ when full information was used) and for the proportion
2:1, there is no detection outside the 99 per cent acceptance
region. Similar qualitative results are found for WF, although with
lower levels of detection for the 1:1 non-Gaussian map (the highest
detection is at the $\sim 7 \sigma$ level). Note that these levels of
detections are slightly below the ones obtained with the LCFC
method. However, some information about the power spectra of the
microwave components is already available and more is expected to be
obtained from the current and future microwave experiments. Therefore
a realistic situation would fall somewhere between the two cases
presented here (full information and no information).

With regard to the LCFC, although the method is in principle less
powerful than MEM and WF, it has the advantage of not requiring so
strong assumptions about the underlying signals. First of all we would
need to have some reasonably good templates for the Galactic
components at each considered frequency channel in order to be able to
subtract the Galactic foreground contamination down to a 10 per
cent. Given that Planck will take observations in a very large range
of frequencies (30 to 857 GHz), including some channels to monitor
Galactic foregrounds, this seems a plausible assumption. In any case,
we have also repeated our analysis for the LCFC without subtracting
any correction for the Galactic foregrounds and the results obtained
were very similar to the ones presented here. Therefore, even if we
were unable to reduce the Galactic contamination by a factor of 10,
the method would still perform similarly well. The second and most
important requirement to construct the LCFC is a good knowledge of the
properties of the noise, which is also a reasonable assumption for the
Planck satellite.

However, the choice of the channels to be combined is an important
issue that must be studied in detail. For the present work, we have
tried many different combinations of data maps and finally chosen to
present our results with the combination of the 143 and 217 GHz
channels. These two channels are relatively clean from Galactic
contamination (which ensures that the amount of spurious
non-Gaussianity introduced in the reconstructed CMB is very limited)
and have a good resolution (important if we are looking for a
non-Gaussian signal present at small scales). From all the tested
combinations that preserved the Gaussian character of the CMB for the
four considered regions of the sky (when the input was Gaussian), this
was the one that yielded the highest non-Gaussian detections (when the
input was non-Gaussian). We would like to remark that the optimal
choice of channels will depend on the area of the sky that is being
analysed as well as on the type of non-Gaussianity that we are trying
to find. For instance, the 353 GHz channel has a good resolution but
it may also have a high contamination from dust emission. However, if
we are studying a particularly clean region of the sky, it can be
helpful to add the 353 GHz channel.  Also, if we are interested in
looking for non-Gaussian signatures at larger scales we could instead
add other channels (e.g. the 100 GHz map). This would reduce the
resolution of the recovered CMB map, but it will increase the
signal-to-noise ratio at the scales of interest.  In addition, the
optimal choice may also depend on the particular test that is used to
study the temperature distribution.



We also have to point out that the proportion of cosmic strings (or
other topological defects) present in the universe, if any, is
uncertain.  According to \cite{bou01}, the best fit of a mixture of
inflation and cosmic strings to the CMB power spectrum obtained from
BOOMERanG (and other CMB experiments) yields to a string contribution
of 18 per cent (in the power spectrum). This corresponds approximately
to a mixture of Gaussian CMB and cosmic strings in proportion 2:1 in
rms.  Using the WMAP and 2dF data, \cite{pog03} conclude that a
cosmic-string contribution $\la 10$ per cent can not be ruled out.  If
the cosmic strings proportion was lower than the cases studied in this
work, the wavelet technique would not be able to detect this
contribution with the reconstructed methods that we have
considered. However, note that we are using small patches of the sky
of size $12.8^\circ \times 12.8^\circ$ for the analysis. But Planck
will provide with all-sky observations of the microwave sky and thus
it is expected that the non-Gaussian analysis can be performed over a
much larger fraction of the sky. Of course having a larger sky area
implies that the proportion of cosmic strings that can be detected is
also appreciably lower. Therefore we expect to obtain qualitatively
similar results to the ones found in this work for non-Gaussian maps
with lower proportions of cosmic strings that cover larger fractions
of the sky.

To understand better the effect of the different contaminants on the
CMB reconstruction, the results in sections~\ref{sec:gaussian}
and~\ref{sec:ng} were obtained from simulated data that did not
include the emission of point sources. In
section~\ref{sec:point_sources} we repeat the non-Gaussianity analysis
including point source emission for an input Gaussian CMB map. In
particular, we perform the CMB reconstruction using MEM and a MEM+MHW
joint analysis. We find that, when using only MEM, point source
residuals contaminate the reconstructed CMB which clearly shows up in
the wavelet statistic. However, this problem is solved when using
the MEM+MHW joint analysis. Therefore using a tool such as the MHW to
remove the brightest point sources prior to applying a component
separation technique is crucial in order to avoid that spurious
non-Gaussianity is present in the reconstructed CMB map. Again, when
considering a larger fraction of the sky, we may wonder if we would be
able to see the point source residuals. As shown in
section~\ref{sec:point_sources_low10}, we do expect to see the
residuals left by this contaminant when analysing a large fraction of
the sky, even after subtracting the brightest point sources. Therefore
it is a critical issue to either remove or mask as much as possible
the contribution coming from this contaminant.

\section{Conclusions}
\label{sec:conclusions}
The aim of this work was to show the importance of
considering the whole component separation process when testing the
performance of statistical tools for Gaussianity analysis. The method used to
reconstruct the CMB map may introduce or erase non-Gaussianity on the
cosmological signal and it is very important to understand this effect in
order to give a correct interpretation of the results of the analysis.
This type of study also allows one to establish which reconstruction
technique is better, not in the sense of recovering the best power
spectrum, but with regard to preserve the underlying CMB temperature
distribution.

In particular, we use a wavelet based technique to test the CMB
recovered by three different methods: MEM, WF and LCFC. For the test
we have used Planck simulated data on small patches of the sky,
considering Gaussian and non-Gaussian CMB maps. Assuming that point
sources have been appropriately removed, none of the methods seem to
include spurious non-Gaussianity. However, when the underlying CMB is
non-Gaussian, the three methods produce CMB reconstructions which are
more Gaussian than the input one. This is mainly due to the presence
of instrumental noise, which significantly affects the CMB
reconstruction, especially at the smallest scales where the non-Gaussian
signatures of our test maps were more important.  MEM tends
to provide slightly larger non-Gaussian detections than WF, whereas
the detections are lower for the LCFC. This is mainly due to the fact
that MEM and WF try to denoise and deconvolve the signal and not so
much to the fact that they remove better the foreground contamination.

When point sources are included, if no attempt to remove the
brightest point sources is done, they clearly contaminate the CMB
reconstructions introducing spurious non-Gaussianity. However, if we
first apply the MHW to remove the brightest point sources, the
reconstructions are again compatible with Gaussianity. This shows the
importance of removing or masking the contamination coming from
extragalactic point sources since otherwise it will impair our ability
to detect intrinsic non-Gaussianity. Moreover, when considering a
larger region of the sky, the point source residuals would be more
statistically significant and we expect to detect them with our
wavelet technique even after subtracting the brightest ones. Therefore
we find that this is the most damaging contaminant for Gaussian
analysis of the CMB, at least at the scales considered in this work
and within the uncertainties of our foreground simulations (that could
be more important at subdegree scales).

\section*{Acknowledgements}
RBB thanks the Ministerio de Ciencia y Tecnolog\'\i a and the
Universidad de Cantabria for a Ram\'on y Cajal contract. 
The authors thank F.R.Bouchet, J.M.Diego and L.Toffolatti for kindly
provide the cosmic strings, SZ and point source simulations, respectively.
We thank the RTN of the EU project HPRN-CT-2000-00124 for partial
financial support. RBB, EMG and PV acknowledge partial financial support
from the Spanish MCYT project ESP2004-07067-C03-01 and from
a joint Spain-France project (Acci\'on Integrada HF03-163, Programmes
d'Actions Int\'egr\'ees (Picasso)).

\end{document}